\documentclass[12 pt,twoside,aps,prd,amsmath,amssymb,
tightenlines,showpacs,showkeys,eqsecnum]{revtex4}
\usepackage{epsfig}
\usepackage{psfig}
\usepackage{mathptmx}
\usepackage{bm}
\usepackage{graphicx}

\usepackage{hyperref}
\renewcommand{\cal}{\mathcal}

\newcommand {\ve}{\varepsilon}

\newcommand {\prm}{\prime}

\newcommand {\cL}{\cal L}
\newcommand {\G}{\Gamma}

\def \myfigures #1#2#3#4#5#6#7#8
{\begin{figure}[ht]
    \begin{center}
        \includegraphics[width=#2 \textwidth, angle=-90]{#1.eps}
        \hfill
        \includegraphics[width=#6 \textwidth, angle=-90]{#5.eps}
        \parbox[t]{#4\textwidth}{\caption {#3}\label{#1}}
        \hfill
        \parbox[t]{#8\textwidth}{\caption {#7}\label{#5}}
    \end{center}
\end{figure} }

\def \myf #1#2#3#4#5#6#7#8
{\begin{figure}[ht]
    \begin{center}
        \includegraphics[width=#1 \textwidth, angle=-90]{#3.eps}
        \hfill
        \includegraphics[width=#1 \textwidth, angle=-90]{#5.eps}
        \hfill
        \includegraphics[width=#1 \textwidth, angle=-90]{#7.eps}
        \parbox[t]{#2\textwidth}{\caption {#4}\label{#3}}
        \hfill
        \parbox[t]{#2\textwidth}{\caption {#6}\label{#5}}
        \hfill
        \parbox[t]{#2\textwidth}{\caption {#8}\label{#7}}
    \end{center}
\end{figure} }

\def\myfigure #1#2#3#4
{\begin{figure}[ht]\begin{center}
\includegraphics[width=#2 \textwidth, angle=-90]{#1.eps}
\parbox[t]{#4\textwidth}{\caption{#3}\label{#1}}
\end{center}\end{figure}}

\begin{document}
\title{Bianchi type I universe with viscous fluid: A qualitative analysis}
\author{Bijan Saha and V. Rikhvitsky}
\affiliation{Laboratory of Information Technologies\\
Joint Institute for Nuclear Research, Dubna\\
141980 Dubna, Moscow region, Russia} \email{saha@thsun1.jinr.ru}
\homepage{http://thsun1.jinr.ru/~saha/}
\date{\today}

\begin{abstract}
The nature of cosmological solutions for a homogeneous,
anisotropic Universe given by a Bianchi type-I (BI) model in the
presence of a Cosmological constant $\Lambda$ is investigated by
taking into account dissipative process due to viscosity. The
system in question is thoroughly studied both analytically and
numerically. It is shown the viscosity, as well as the $\Lambda$
term exhibit essential influence on the character of the
solutions. In particular a negative $\Lambda$ gives rise to an
ever-expanding Universe, whereas, a suitable choice of initial
conditions plus a positive $\Lambda$ can result in a
singularity-free oscillatory mode of expansion. For some special
cases it is possible to obtain oscillations in the exponential
mode of expansion of the BI model even with a negative $\Lambda$,
where oscillations arise by virtue of viscosity.
\end{abstract}



\maketitle

\bigskip


\section{Introduction}

The investigation of relativistic cosmological models usually has
the energy momentum tensor of matter generated by a perfect fluid.
To consider more realistic models one must take into account the
viscosity mechanisms, which have already attracted the attention
of many researchers. Misner \cite{mis1,mis2} suggested that strong
dissipative due to the neutrino viscosity may considerably reduce
the anisotropy of the blackbody radiation. Viscosity mechanism in
cosmology can explain the anomalously high entropy per baryon in
the present universe \cite{wein,weinb}. Bulk viscosity associated
with the grand-unified-theory phase transition \cite{lang} may
lead to an inflationary scenario \cite{waga,pacher,guth}.

A uniform cosmological model filled with fluid which possesses pressure
and second (bulk) viscosity was developed by Murphy \cite{murphy}. The
solutions that he found exhibit an interesting feature that the big bang
type singularity appears in the infinite past. Exact solutions of the
isotropic homogeneous cosmology for open, closed and flat universe have
been found by Santos et al \cite{santos}, with the bulk viscosity being
a power function of energy density.

The nature of cosmological solutions for homogeneous Bianchi type
I (BI) model was investigated by Belinsky and Khalatnikov
\cite{belin} by taking into account dissipative process due to
viscosity. They showed that viscosity cannot remove the
cosmological singularity but results in a qualitatively new
behavior of the solutions near singularity. They found the
remarkable property that during the time of the \textit{big bang}
matter is created by the gravitational field. BI solutions in case
of stiff matter with a shear viscosity being the power function of
of energy density were obtained by Banerjee \cite{baner}, whereas
BI models with bulk viscosity ($\eta$) that is a power function of
energy density $\ve$ and when the universe is filled with stiff
matter were studied by Huang \cite{huang}. The effect of bulk
viscosity, with a time varying bulk viscous coefficient, on the
evolution of isotropic FRW models was investigated in the context
of open thermodynamics system was studied by Desikan
\cite{desikan}. This study was further developed by Krori and
Mukherjee \cite{krori} for anisotropic Bianchi models.
Cosmological solutions with nonlinear bulk viscosity were obtained
in \cite{chim}. Models with both shear and bulk viscosity were
investigated in \cite{elst,meln}.

Though Murphy \cite{murphy} claimed that the introduction of bulk
viscosity can avoid the initial singularity at finite past,
results obtained in \cite{barrow} show that, it is, in general,
not valid, since for some cases big bang singularity occurs in
finite past.

We studied a self-consistent system of the nonlinear spinor and/or
scalar fields in a BI spacetime  in presence of a perfect fluid
and a $\Lambda$ term \cite{sahaprd,bited} in order to clarify
whether the presence of a singular point an inherent property of
the relativistic cosmological models or is it only a consequence
of specific simplifying assumptions underlying these models?
Recently we have considered a system of nonlinear spinor field in
a BI Universe filled with viscous fluid \cite{romrep}. Since the
viscous fluid itself presents a growing interest, we have studied
the influence of viscous fluid and $\Lambda$ term in the evolution
of the BI Universe \cite{sahaprd1}. In that paper we consider only
some special cases those allow exact solutions. In this paper
along with those special cases we study some general cases, giving
a qualitative analysis of the system of equations. We also perform
some numerical calculations and compare the results obtained with
those given in some pioneering papers in this field, e.g.
\cite{belin}

        \section{Derivation of Basic Equations}
Using the variational principle in this section we derive the
fundamental equations for the gravitational field  from the action
\eqref{action} :
\begin{equation}
{\cal S}(g; \ve) = \int\, \cL \sqrt{-g} d\Omega \label{action}
\end{equation}
with
\begin{equation}
\cL= \cL_{\rm grav.} + \cL_{\rm vf}. \label{lag}
\end{equation}

The gravitational part of the Lagrangian \eqref{lag} ${\cL}_{\rm
grav.}$ is given by a Bianchi type-I metric, whereas the term
${\cL}_{\rm vf}$ describes a viscous fluid.

We also write the expressions for the metric functions explicitly
in terms of the volume scale $\tau$ defined bellow \eqref{taudef}.
Defining Hubble constant \eqref{hubc} in analogy with a flat
Friedmann-Robertson-Walker (FRW) Universe, we also derive the
system of equations for $\tau$, $H$ and $\ve$, with $\ve$ being
the energy density of the viscous fluid, which plays the central
role here.

             \subsection{The gravitational field}
As a gravitational field we consider the Bianchi type I (BI) cosmological
model. It is the simplest model of anisotropic universe that describes
a homogeneous and spatially flat space-time and if filled with perfect
fluid with the equation of state $p = \zeta \ve, \quad \zeta < 1$, it
eventually evolves into a FRW universe \cite{jacobs}. The isotropy of
present-day universe makes BI model a prime candidate for studying the
possible effects of an anisotropy in the early universe on modern-day
data observations. In view of what has been mentioned above we choose
the gravitational part of the Lagrangian \eqref{lag} in the form
\begin{equation}
{\cal L}_{\rm g} = \frac{R}{2\kappa},
\label{lgrav}
\end{equation}
where $R$ is the scalar curvature, $\kappa = 8 \pi G$
being the Einstein's gravitational constant. The gravitational field in
our case is given by a Bianchi type I (BI) metric
\begin{equation}
ds^2 = dt^2 - a^2 dx^2 - b^2 dy^2 - c^2 dz^2,
\label{BI}
\end{equation}
with $a,\, b,\, c$ being the functions of time $t$ only. Here the speed of
light is taken to be unity.

\subsection{Viscous fluid}

The influence of the viscous fluid in the evolution of the
Universe is performed by means of its energy momentum tensor,
which acts as the source of the corresponding gravitational field.
The reason for writing $\cL_{\rm vf}$ in \eqref{lag} is to
underline that we are dealing with a self-consistent system. The
energy momentum tensor of a viscous field has the form
\begin{equation}
T_{\mu\,{\rm (m)}}^{\nu} = (\ve + p^\prm) u_\mu u^\nu - p^{\prm}
\delta_\mu^\nu + \eta g^{\nu \beta} [u_{\mu;\beta}+u_{\beta:\mu}
-u_\mu u^\alpha u_{\beta;\alpha} - u_\beta u^\alpha
u_{\mu;\alpha}], \label{imper}
\end{equation}
where
\begin{equation}
p^{\prm} = p - (\xi - \frac{2}{3} \eta) u^\mu_{;\mu}. \label{ppr}
\end{equation}
Here $\ve$ is the energy density, $p$ - pressure, $\eta$ and $\xi$
are the coefficients of shear and bulk viscosity, respectively.
Note that the bulk and shear viscosities, $\eta$ and $\xi$, are
both positively definite, i.e.,
\begin{equation}
\eta > 0, \quad \xi > 0.
\end{equation}
They may be either constant or function of time or energy, such as
:
\begin{equation}
\eta = |A| \ve^{\alpha}, \quad \xi = |B| \ve^{\beta}.
\label{etaxi}
\end{equation}
The pressure $p$ is connected to the energy density by means of a
equation of state. In this report we consider the one describing a
perfect fluid :
\begin{equation}
p = \zeta \ve, \quad \zeta \in (0, 1]. \label{pzeta}
\end{equation}
Note that here $\zeta \ne 0$, since for dust pressure, hence
temperature is zero, that results in vanishing viscosity.

In a comoving system of reference such that $u^\mu =
(1,\,0,\,0,\,0)$ we have
\begin{subequations}
\label{total}
\begin{eqnarray}
T_{0\,{\rm (m)}}^{0} &=& \ve, \\
T_{1\,{\rm (m)}}^{1} &=& - p^{\prm} + 2 \eta \frac{\dot a}{a}, \\
T_{2\,{\rm (m)}}^{2} &=& - p^{\prm} + 2 \eta \frac{\dot b}{b}, \\
T_{3\,{\rm (m)}}^{3} &=& - p^{\prm} + 2 \eta \frac{\dot c}{c}.
\end{eqnarray}
\end{subequations}

Let us introduce the dynamical scalars such as the expansion and
the shear scalar as usual
\begin{equation}
\theta = u^\mu_{;\mu},\quad \sigma^2 = \frac{1}{2} \sigma_{\mu\nu}
\sigma^{\mu\nu}, \label{dynscal}
\end{equation}
where
\begin{equation}
\sigma_{\mu\nu} = \frac{1}{2} \Bigl(u_{\mu;\alpha}
P^{\alpha}_{\nu} + u_{\nu;\alpha} P^{\alpha}_{\mu}\Bigr) -
\frac{1}{3} \theta P_{\mu \nu}. \label{shearten}
\end{equation}
Here $P$ is the projection operator obeying
\begin{equation}
P^2 = P.
\end{equation}
For the space-time with signature $(+, \,-,\,-,\,-)$ it has the
form
\begin{equation}
P_{\mu\nu} = g_{\mu\nu} - u_\mu u_\nu, \quad P^\mu_\nu =
\delta^\mu_\nu - u^\mu u_\nu. \label{projec}
\end{equation}
For the BI metric the dynamical scalar has the form
\begin{equation}
\theta = \frac{\dot {a}}{a}+\frac{\dot {b}}{b} + \frac{\dot
{c}}{c} = \frac{\dot {\tau}}{\tau}, \label{expan}
\end{equation}
and
\begin{equation}
2 \sigma^2 = \frac{\dot {a}^2}{a^2}+\frac{\dot {b}^2}{b^2} +
\frac{\dot {c}^2}{c^2} - \frac{1}{3} \theta^2. \label{shesc}
\end{equation}

\subsection{Field equations and their solutions}

Variation of \eqref{action} with respect to metric tensor
$g_{\mu\nu}$ gives the Einstein's field equation. In account of
the $\Lambda$-term for the BI space-time \eqref{BI} this system of
equations can be rewritten as
\begin{subequations}
\label{BID}
\begin{eqnarray}
\frac{\ddot b}{b} +\frac{\ddot c}{c} + \frac{\dot b}{b}\frac{\dot
c}{c}&=&  \kappa T_{1}^{1} -\Lambda,\label{11}\\
\frac{\ddot c}{c} +\frac{\ddot a}{a} + \frac{\dot c}{c}\frac{\dot
a}{a}&=&  \kappa T_{2}^{2} - \Lambda,\label{22}\\
\frac{\ddot a}{a} +\frac{\ddot b}{b} + \frac{\dot a}{a}\frac{\dot
b}{b}&=&  \kappa T_{3}^{3} - \Lambda,\label{33}\\
\frac{\dot a}{a}\frac{\dot b}{b} +\frac{\dot b}{b}\frac{\dot c}{c}
+\frac{\dot c}{c}\frac{\dot a}{a}&=&  \kappa T_{0}^{0} - \Lambda,
\label{00}
\end{eqnarray}
\end{subequations}
where over dot means differentiation with respect to $t$ and
$T_{\nu}^{\mu}$ is the energy-momentum tensor of a viscous fluid
given above \eqref{total}.

\subsubsection{Expressions for the metric functions }

To write the metric functions explicitly, we define a new time
dependent function $\tau (t)$
\begin{equation}
\tau = a b c = \sqrt{-g},
\label{taudef}
\end{equation}
which is indeed the volume scale of the BI space-time.

Let us now solve the Einstein equations. In account of
\eqref{total} from \eqref{11}, \eqref{22}, and \eqref{33} one
finds the following expressions for the metric functions
explicitly~\cite{sahaprd1}
\begin{subequations}
\label{abc}
\begin{eqnarray}
a(t) &=& A_{1} \tau^{1/3}\exp\biggl[(B_1/3) \int\, \frac{ e^{-2
\kappa \int \eta dt}}{\tau}\,dt \biggr], \label{a} \\
b(t) &=& A_{2} \tau^{1/3}\exp\biggl[(B_2/3) \int\, \frac{ e^{-2
\kappa \int \eta dt}}{\tau}\,dt \biggr], \label{b}\\
c(t) &=& A_{3} \tau^{1/3}\exp\biggl[(B_3/3) \int\, \frac{ e^{-2
\kappa \int \eta dt}}{\tau}\,dt\biggr], \label{c}
\end{eqnarray}
\end{subequations}
where the constants $A_i$'s and $B_i$'s obey the following
relations
\begin{eqnarray}
A_1 A_2 A_3 &=& 1, \nonumber\\
B_1 + B_2 + B_3  &=& 0. \nonumber
\end{eqnarray}
Thus, the metric functions are found explicitly in terms of $\tau$
and viscosity.

As one sees from \eqref{a}, \eqref{b} and \eqref{c}, for $\tau = t^n$
with $n > 1$ the exponent tends to unity at large $t$, and the
anisotropic model becomes isotropic one.

\subsubsection{Singularity analysis}

Let us now investigate the existence of singularity (singular
point) of the gravitational case, which can be done by
investigating the invariant characteristics of the space-time. In
general relativity these invariants are composed from the
curvature tensor and the metric one. In a 4D Riemann space-time
there are 14 independent invariants. Instead of analyzing all 14
invariants, one can confine this study only in 3, namely the
scalar curvature $I_1 = R$, $I_2 = R_{\mu\nu}R^{\mu\nu}$, and the
Kretschmann scalar $I_3 =
R_{\alpha\beta\mu\nu}R^{\alpha\beta\mu\nu}$ \cite{bronshik,fay}..
At any regular space-time point, these three invariants
$I_1,\,I_2,\,I_3$ should be finite. Let us rewrite these
invariants in detail.

For the Bianchi I metric one finds the scalar curvature
\begin{eqnarray}
I_1 = R = - 2\Bigl(\frac{\ddot a}{a}+\frac{\ddot b}{b}+\frac{\ddot c}{c}+
\frac{\dot a}{a}\frac{\dot b}{b} + \frac{\dot b}{b}\frac{\dot c}{c}+
\frac{\dot c}{c}\frac{\dot a}{a}\Bigr).
\label{SC}
\end{eqnarray}
Since the Ricci tensor for the BI metric is diagonal, the
invariant $I_2 = R_{\mu\nu}R^{\mu\nu} \equiv R_{\mu}^{\nu} R_{\nu}^{\mu}$
is a sum of squares of diagonal components of Ricci tensor, i.e.,
\begin{equation}
I_2 = \Bigl[\bigl(R_{0}^{0}\bigr)^2  + \bigl(R_{1}^{1}\bigr)^2 +
\bigl(R_{2}^{2}\bigr)^2 + \bigl(R_{3}^{3}\bigr)^2 \Bigr].
\end{equation}

Analogically, for the Kretschmann scalar in this case we have $I_3
= R_{\,\,\,\,\,\,\,\alpha\beta}^{\mu\nu}
R_{\,\,\,\,\,\,\,\mu\nu}^{\alpha\beta}$, a sum of squared
components of all nontrivial $R_{\,\,\,\,\,\,\,\mu\nu}^{\mu\nu}$,
which can be written as
\begin{eqnarray}
I_3 &=& 4 \Biggl[
 \Bigl(R_{\,\,\,\,\,\,01}^{01}\Bigr)^2
+ \Bigl(R_{\,\,\,\,\,\,02}^{02}\Bigr)^2
+ \Bigl(R_{\,\,\,\,\,\,03}^{03}\Bigr)^2
+ \Bigl(R_{\,\,\,\,\,\,12}^{12}\Bigr)^2
+ \Bigl(R_{\,\,\,\,\,\,23}^{23}\Bigr)^2
+ \Bigl(R_{\,\,\,\,\,\,31}^{31}\Bigr)^2\Biggr] \nonumber\\
&=& 4\Bigl[\Bigl(\frac{\ddot a}{a}\Bigr)^2 +
\Bigl(\frac{\ddot b}{b}\Bigr)^2+\Bigl(\frac{\ddot c}{c}\Bigr)^2
+ \Bigl(\frac{\dot a}{a}\frac{\dot b}{b}\Bigr)^2 +
\Bigl(\frac{\dot b}{b}\frac{\dot c}{c}\Bigr)^2 +
\Bigl(\frac{\dot c}{c}\frac{\dot a}{a}\Bigr)^2\Bigr].
\label{Kretsch}
\end{eqnarray}
Let us now express the foregoing invariants in terms of $\tau$.
From Eqs. \eqref{abc} we have
\begin{subequations}
\label{sing}
\begin{eqnarray}
a_i &=& A_i \tau^{1/3} \exp \Biggl((B_i/3) \int\,\frac{e^{-2
\kappa \int \eta dt}}{\tau (t)} dt\Biggr),
\\
\frac{{\dot a}_i}{a_i} &=& \frac{\dot{\tau} + B_i e^{-2 \kappa
\int \eta dt}}{3 \tau}
\quad (i=1,2,3,),\\
\frac{{\ddot a}_i}{a_i} &=& \frac{3 \tau \ddot{\tau} - 2
\dot{\tau}^2 - \dot{\tau} B_i e^{-2 \kappa \int \eta dt} - 6
\kappa \eta \tau B_i e^{-2 \kappa \int \eta dt} + B_i^2 e^{-4
\kappa \int \eta dt}} {9 \tau^2},
\end{eqnarray}
\end{subequations}
i.e., the metric functions $a, b, c$ and their derivatives are in
functional dependence with $\tau$. From Eqs. \eqref{sing} one can
easily verify that \cite{sahaprd1}
$$I_1 \propto \frac{1}{\tau^2},\quad
I_2 \propto \frac{1}{\tau^4},\quad I_3 \propto \frac{1}{\tau^4}.$$
Thus we see that at any space-time point, where $\tau = 0$ the
invariants $I_1,\,I_2,$ and $I_3$ become infinity, hence the
space-time becomes singular at this point.

\subsection{Equations for determining $\tau$}

In the foregoing subsection we wrote the corresponding metric
functions in terms of volume scale $\tau$. In what follows, we
write the equation for $\tau$ and study it in details.

Summation of Einstein equations \eqref{11}, \eqref{22}, \eqref{33}
and 3 times \eqref{00} gives
\begin{equation}
{\ddot \tau} - \frac{3}{2} \kappa \xi {\dot \tau} =
\frac{3}{2}\kappa \bigl(\ve - p \bigr) \tau - 3 \Lambda \tau.
\label{dtau1}
\end{equation}
For the right-hand-side of \eqref{dtau1} to be a function of
$\tau$ only, the solution to this equation is
well-known~\cite{kamke}.

The energy-momentum conservation law, i.e.,
\begin{equation}
T_{\mu;\nu}^{\nu} = T_{\mu,\nu}^{\nu} + \G_{\rho\nu}^{\nu}
T_{\mu}^{\rho} - \G_{\mu\nu}^{\rho} T_{\rho}^{\nu} = 0,
\end{equation}
in our case gives the following equation for $\ve$ :
\begin{equation}
{\dot \ve} + \frac{\dot \tau}{\tau} \omega - (\xi + \frac{4}{3}
\eta) \frac{{\dot \tau}^2}{\tau^2} + 4 \eta (\kappa T_0^0 -
\Lambda) = 0, \label{vep}
\end{equation}
where
\begin{equation}
\omega = \ve + p, \label{thermal}
\end{equation}
is the thermal function.

Defining a generalized Hubble constant $H$ :
\begin{equation}
\frac{\dot {\tau}}{\tau} = \frac{\dot {a}}{a}+\frac{\dot {b}}{b} +
\frac{\dot {c}}{c} = 3 H. \label{hubc}
\end{equation}
the Eqs. \eqref{dtau1} and \eqref{vep} in account of \eqref{total}
can be rewritten as
\begin{subequations}
\label{HVe}
\begin{eqnarray}
\dot {H} &=& \frac{\kappa}{2}\bigl(3 \xi H - \omega\bigr) -
\bigl(3 H^2 - \kappa \ve + \Lambda \bigr) ,  \label{H}\\
\dot {\ve} &=& 3 H\bigl(3 \xi H - \omega\bigr) + 4 \eta \bigl(3
H^2 - \kappa \ve + \Lambda \bigr) . \label{Ve}
\end{eqnarray}
\end{subequations}
In terms of dynamical scalars $\theta$ and $\sigma$ the system
\eqref{HVe} takes a very simple form
\begin{subequations}
\label{TS}
\begin{eqnarray}
\dot {\theta} &=& \frac{3\kappa}{2}\bigl(\xi \theta - \omega\bigr)
- 3 \sigma^2,  \label{theta}\\
\dot {\ve} &=& \theta \bigl(\xi \theta - \omega\bigr) + 4 \eta
\sigma^2. \label{Vsig}
\end{eqnarray}
\end{subequations}
Note that the Eqs. \eqref{TS} coincide with the ones given in
\cite{baner}.

 \section{Qualitative analysis and some special solutions}

In this subsection we simultaneously solve the system of equations
for $\tau$, $H$, and $\ve$. It is convenient to rewrite the Eqs.
\eqref{hubc} and \eqref{HVe} as a single system :
\begin{subequations}
\label{HVen}
\begin{eqnarray}
\dot \tau &=& 3 H \tau, \label{tauH}\\
\dot {H} &=& \frac{\kappa}{2}\bigl(3 \xi H - \omega\bigr) -
\bigl(3 H^2 - \kappa \ve + \Lambda \bigr) ,  \label{Hn}\\
\dot {\ve} &=& 3 H\bigl(3 \xi H - \omega\bigr) + 4 \eta \bigl(3
H^2 - \kappa \ve + \Lambda \bigr) . \label{Ven}
\end{eqnarray}
\end{subequations}

In account of \eqref{thermal},\eqref{etaxi} and \eqref{pzeta} the
Eqs. \eqref{HVen} now can be rewritten as
\begin{subequations}
\label{HVen1}
\begin{eqnarray}
\dot \tau &=& 3 H \tau, \label{tauH1}\\
\dot {H} &=& \frac{\kappa}{2}\bigl(3 B \ve^\beta H - (1+\zeta)\ve
\bigr) - \bigl(3 H^2 - \kappa\ve + \Lambda \bigr) ,
\label{Hn1}\\
\dot {\ve} &=& 3 H\bigl(3 B \ve^\beta H - (1+\zeta)\ve \bigr) + 4
A \ve^\alpha \bigl(3 H^2 -  \kappa \ve + \Lambda \bigr) .
\label{Ven1}
\end{eqnarray}
\end{subequations}

The system \eqref{HVen} have been extensively studied in
literature either partially \cite{murphy,huang,baner} or in
general \cite{belin}. In what follows, we consider the system
\eqref{HVen} for some special choices of the parameters.

\subsection{Qualitative analysis}

Following Belinski and Khalatnikov \cite{belin} let us now study
the characters of the solutions of the dynamical system
\eqref{HVen} or \eqref{HVen1}. We first rewrite the system
\eqref{HVen}, namely \eqref{Hn} and \eqref{Ven} in the matrix form
:
\begin{equation}
\left(
\begin{array}{c}
\dot {H}\\
\dot{\ve}
\end{array}
\right) = \left( \begin{array}{ccc} \kappa/2& & -1\\
3 H &  &4 \eta\end{array}\right) \left(
\begin{array}{c}
3 \xi H - \omega\\
3 H^2 - \kappa \ve + \Lambda
\end{array}
\right). \label{eqmatr}
\end{equation}
Note that unlike the system studied by Belinski and Khalatnikov
the system in consideration contains a Cosmological constant
$\Lambda$.

\subsubsection{General properties of the system}

Easy to note that the solutions cannot intersect the axis $\ve =
0$, since $\dot \ve |_{\ve = 0} = 0$, as well as the parabola
\begin{equation}
3 H^2 - \kappa \ve + \Lambda = 0, \label{parab1}
\end{equation}
as far as \eqref{parab1} is itself the integral curve. Thus,
starting from the point $(H, \ve) = (+ \infty, 0)$, the solutions
cannot enter into the "prohibited region" inside the parabola
\eqref{parab1}. Whether they may achieve $H < 0$ depends on the
value of $\Lambda$. Note that, unlike the system considered by
Belinski {\it et. al} \cite{belin} the system in this report
contains a nonzero $\Lambda$ term.

\subsubsection{Critical points of the dynamical system}

{\bf a)} By virtu of linear independence of the columns of the
matrix of the Eq. \eqref{eqmatr} the critical points are the
solutions of the equations
\begin{subequations}
\label{cheq}
\begin{eqnarray}
3 \xi H - \omega &=& 0, \label{ch1}\\
3 H^2 - \kappa \ve + \Lambda &=& 0. \label{ch2}
\end{eqnarray}
\end{subequations}
i.e., they necessarily lie on the parabola \eqref{parab1}.
Solutions to the system \eqref{cheq} will be the roots of the
equation
\begin{subequations}
\label{C}
\begin{eqnarray}
 3 \kappa B^2 \ve^{1 + 2 \beta} - (1 + \zeta)^2 \ve^2 &-& 3
\Lambda B^2 \ve^{2 \beta} = 0,\\
H &=& \frac{1 + \zeta}{3 B} \ve^{1-\beta}.
\end{eqnarray}
\end{subequations}

The quantity of the positive roots of the Eq. \eqref{C} according
to Cartesian law is equal to the number of changes of sign of
coefficients of equations or less than that by an even number. So,
for \begin{eqnarray}
\begin{array}{ccccc}
 & \Lambda < 0 & \,\,\, {\rm and} \,\,\, & 1/2 < \beta < 1 &
(Fig. \ref{he2}, Fig. \ref{he3}) \nonumber \\
or & \,\,\, \Lambda > 0 & {\rm and} & \beta < 1/2 & (Fig.
\ref{he5}, Fig. \ref{he6})
\end{array}
\end{eqnarray}
the number of roots is either $2$ or zero. For the remaining cases
\begin{eqnarray}
\begin{array}{ccccc}
 & \Lambda < 0 & \,\,\, {\rm and} \,\,\, & \beta > 1 & (Fig.\,
 \ref{he1}), \nonumber \\
 & \Lambda < 0 & {\rm and} & \beta < 1/2 & (Fig.\,
 \ref{he4}), \nonumber \\
 & \Lambda > 0 & {\rm and} & \beta > 1/2 & (Fig.\,
 \ref{he7}) \nonumber
\end{array}
\end{eqnarray}
there exists only one root. The corresponding pictures of the
phase curves are given in figures cited above. The critical points
are denoted by small circles. Note that here we consider the case
with $\eta = 0$, i.e., $A = 0$. In case if $\eta \ne 0$, with the
increase of $A$ the separatrix of the saddle tilts (inclines) to
the left. Since the overall picture for $A \ne 0$ remains
qualitatively unaltered, we only show the corresponding phase
portrait for two cases, namely Fig. \ref{he8} corresponds to Fig.
\ref{he1}, Fig. \ref{he9} corresponds to Fig. \ref{he4}. Note that
for numerical calculations we set $\kappa = 1$, $\zeta = 0.333$
(if not mentioned otherwise). In the Figs. \ref{he1} - \ref{he7}
$\eta$ is taken to be zero. Note that in the Figs. $E$ and $T$
stand for $\ve$ and $\tau$, respectively.

Since, the equation for $\ve$ only contains $\eta$, the energy
density for nontrivial $\eta$ undergoes essential changes, whereas
$H$ and $\tau$ remain virtually unchanged.

The types of critical points lying on the integral curve
alternate: $\ldots$ saddle, attracting knot, saddle $\ldots$. So
it is sufficient to consider the case with maximum number of
roots. Taking into account the Eqs. \eqref{Ven} and \eqref{parab1}
let us now calculate
\begin{eqnarray}
\lim_{\ve \to +\infty} \frac{\dot \ve}{3 H \ve} &=& \lim_{\ve \to
+\infty} \frac{3 B \ve^\beta \sqrt{\kappa \ve - \Lambda} - \ve (1
+ \zeta)}{\ve} \nonumber \\
&=& 3 B \sqrt{\kappa \ve^{(2\beta -1)} - \Lambda \ve^{-2}} - (1 +
\zeta) = \left\{\begin{array}{ccccc}
-(1 + \zeta) & < & 0,& \beta < 1/2, \\
\\
+ \infty & > & 0, & \beta > 1/2. \\
\end{array}\right.
\end{eqnarray}
So, the latest critical point for $\beta < 1/2$ is attracting knot
and for $\beta > 1/2$ is saddle.

{\bf b)}  It is obvious that if $\,\,\,Lambda \leq 0\,\,\,$ the
points of intersection of the boundary are the critical points
\begin{subequations}
\begin{eqnarray}
H &=& \pm \sqrt{-\Lambda/3},\\
\ve &=& 0.
\end{eqnarray}
\end{subequations}

{\bf c)} For $H < 0$ there may exist critical points , if the
columns of the matrix of \eqref{eqmatr} are linearly dependent. In
that case the critical points are the roots of the equation
\begin{equation}
3 \kappa (\zeta -1)\ve + 6 \kappa^2 A B \ve^{\alpha + \beta}+ 8
\kappa^2 A^2 \ve^{2 \alpha} + 6 \Lambda  = 0,
\end{equation}
and
\begin{equation}
H = -\frac{2}{3} \kappa A \ve^{\alpha}.
\end{equation}

In case of $\eta = 0$ the roots of the characteristic equation
\begin{equation}
\Bigl|\frac{D({\dot H},\,{\dot \ve})}{D(H,\,\ve)} - \mu \Bigr| =
0,
\end{equation}
are
\begin{equation}
\mu_{1,2} = \frac{3 \kappa \xi \pm \sqrt{9 \kappa^2 \xi^2 - 48
\Lambda (1 + \zeta)}}{4}.
\end{equation}
The critical point $\,\,\,(H,\,\ve) =
(0,\,2\Lambda/[\kappa(1-\zeta)])\,\,\,$ is of type divergent focus
if $\,\,\Lambda > 9 \kappa^2 \xi^2/[48(1 + \zeta)]\,\,$ or
divergent knot if $\,\,\Lambda < 9 \kappa^2 \xi^2/[48(1 +
\zeta)]\,\,$.

In the cases illustrated in Figs. \ref{he5} and \ref{he7}, $H \to
\infty$ and $\ve \to \infty$ as $t \to \infty$, whereas, for the
cases given in Fig. \ref{he6} one observes increasing oscillation
bounded by the attracting parabola \eqref{parab1}.

\subsubsection{Integral curves}

For $\Lambda \le 0$ the solutions starting from the upper
half-plane $H > 0$ cannot enter into the lower one. For $\Lambda >
0$ some of the solutions may enter into the lower half-plane
through the segment $H = 0$ and $0 \leq \ve \leq \Lambda$ and
never returns back, since $\dot{H}|_{H=0} < 0.$

\subsection{Numerical solutions}

In this subsection solutions to the system of equations
\eqref{HVen} has been obtained numerically. Evolution of the
Hubble constant $H$, energy density $\ve$ and volume scale $\tau$
corresponding to the cases studied above with different $B$,
$\beta$ and $\Lambda$ has been illustrated in the Figs. \ref{h1} -
\ref{t7}. As one sees, for a negative $\Lambda$ the volume scale
$\tau$ expands exponentially, whereas, for a positive $\Lambda$
there exist solutions where $\tau$ initially expands and after
reaching some maximum begins to contract and finally collapses
into a point, thus giving rise to space-time singularity. Beside
this, as one sees from Fig. \ref{t10}, a suitable choice of
initial conditions gives rise to a singularity-free oscillatory
mode of expansion of the Universe.

\subsection{exact solutions}

In this subsection we consider some special cases allowing exact
solutions.

        \subsubsection{Case with bulk viscosity}
Let us first consider the case when the real fluid possesses the
bulk viscosity only. The corresponding system of Eqs. can then be
obtained by setting $\eta = 0$ in \eqref{HVen} or $A = 0$ in
\eqref{HVen1}. In this case the Eqs. \eqref{tauH} and \eqref{Hn}
remain unaltered, while \eqref{Ven} takes the form
\begin{equation}
\dot {\ve} = 3 H\bigl(3 \xi H - \omega\bigr). \label{vetau}
\end{equation}
In view of \eqref{vetau} the system \eqref{HVen} admits the
following first integral
\begin{equation}
\tau^2 \bigl(\kappa \ve - 3 H^2 - \Lambda \bigr) = C_1, \quad C_1
= {\rm const.} \label{int1}
\end{equation}
The relation \eqref{int1} can be interpreted as follows. At the
initial stage of evolution the volume scale $\tau$ tends to zero,
while, the energy density $\ve$ tends to infinity. Since the
Hubble constant and the $\Lambda$ term are finite, the relation
\eqref{int1} is in correspondence with the current line of
thinking. Let us see what happens as the Universe expands. It is
well known that with the expansion of the Universe, i.e., with the
increase of $\tau$, the energy density $\ve$ decreases. Suppose at
some stage of expansion $\tau \to \infty$, hence $\ve \to 0$. Then
from \eqref{int1} follows that at the stage in question
\begin{equation}
3 H^2 + \Lambda \to 0. \label{asympan}
\end{equation}
In case of $\Lambda = 0$, we find $H = 0$, i.e., in absence of a
$\Lambda$ term, once $\tau \to \infty$, the process of evolution
is terminated. As one sees from \eqref{asympan}, for the $H$ to
make any sense, the $\Lambda$ term should be negative. In presence
of a negative $\lambda$ term the evolution process of the Universe
never comes to a halt, it either expands further or begin to
contract depending on the sign of $H = \pm \sqrt{- \Lambda/3}$,
\quad $\Lambda < 0$.

Let us now consider the case when the bulk viscosity is inverse
proportional to expansion, i.e.,
\begin{equation}
\xi \theta = C_2, \quad C_2 = {\rm const.} \label{xthinv}
\end{equation}
Now keeping into mind that $\theta = {\dot \tau}/\tau = 3 H$, also
the relations \eqref{tauH}, \eqref{thermal} and \eqref{pzeta} the
Eq. \eqref{vetau} can be written as
\begin{equation}
\frac{\dot \ve}{C_2 - (1 + \zeta) \ve} = \frac{\dot \tau}{\tau}.
\label{vetau1}
\end{equation}
From the Eq. \eqref{vetau1} one finds
\begin{equation}
\ve = \frac{1}{1 + \zeta} \bigl[C_2 + C_3 \tau^{-(1 +
\zeta)}\bigr], \label{vetauex}
\end{equation}
with $C_3$ being some arbitrary constant. Further, inserting $\ve$
from \eqref{vetauex} into \eqref{dtau1} one finds the expression
for $\tau$ explicitly.

Taking into account the equation of state \eqref{pzeta} in view of
\eqref{xthinv} and \eqref{vetauex}, the Eq. \eqref{dtau1} admits
the following solution in quadrature :
\begin{equation}
\int \frac{d \tau}{\sqrt{C_2^2 + C_0^0 \tau^2 + C_1^1 \tau^{1 -
\zeta}}} = t + t_0, \label{quadr}
\end{equation}
where $C_2^2$ and $t_0$ are some constants. Further we set $t_0 =
0$. Here, $C_0^0 = 3 \kappa C_2/(1 + \zeta) - 3 \Lambda$ and
$C_1^1 = 3 \kappa C_3 /(1 + \zeta)$. As one sees, $C_0^0$ is
negative for
\begin{equation}
\Lambda > \kappa C_2/(1 + \zeta). \label{Lamp}
\end{equation}
It means that for a positive $\Lambda$ obeying \eqref{Lamp} (we
assume that the constant $C_2$ is a positive quantity) $\tau$
should be bound from above as well. It should be noted that for a
suitable choice of $C_2^2$ and $\tau_0$ (the initial value of
$\tau$), it is possible to obtain oscillatory mode of expansion
with $\tau$ being always positive, i.e., a singularity free
evolution of the Universe. The phase portrait of the $(H,\,\ve)$
plane and the evolution of the BI Universe corresponding to this
portrait allowing oscillatory solutions are given in Figs.
\ref{he10} and \ref{t10}.

As a second example we consider the case, when $\zeta = 1$. From
\eqref{quadr} one then finds
\begin{subequations}
\label{exposc}
\begin{eqnarray}
\tau (t) &=& \bigl(\exp(\sqrt{C_0^0}\,\,t) - C_2^2
\exp(\sqrt{C_0^0}\,\,t)\bigr)/(2 \sqrt{C_0^0}), \quad C_0^0 > 0, \label{exp} \\
\tau (t) &=& (C_2^2 / \sqrt{|C_0^0|}) \sin\,(\sqrt{|C_0^0|}\,\,
t). \quad C_0^0 < 0 \label{osc}.
\end{eqnarray}
\end{subequations}
Taking into account that $C_0^0 > 0$ for any non-positive
$\Lambda$, from \eqref{exp} one sees that, in case of $\Lambda \le
0$ the Universe may be infinitely large (there is no upper bound),
which is in line with the conclusion made above. On the other
hand, $C_0^0$ may be negative only for some positive value of
$\Lambda$. Thus we see that a positive $\Lambda$ can generate a
oscillatory mode of expansion of a BI Universe. The oscillation
takes place around the critical point $(H,\,\ve) = (0,\,(2\Lambda
-\kappa C_2)/[\kappa (1 - \zeta)])$ having the type of cycle under
the condition $\Lambda > \kappa C_2/(1 + \zeta)$. It was shown in
\cite{sahaprd,bited} that in case of a perfect fluid a positive
$\Lambda$ always invokes oscillations in the model, whereas, in
the present model with viscous fluid, it is the case only when
$\Lambda$ obeys \eqref{Lamp}. Unlike the case with radiation where
BI admits a singularity-free oscillatory mode of evolution, here,
in case of a stiff matter one finds the BI Universe first expands,
reaches its maximum and then contracts into a point, thus giving
rise to space-time singularity.

\subsubsection{Case with shear and bulk viscosity}

Let us now consider the general case with the shear viscosity
$\eta$ being proportional to the expansion, i.e.,
\begin{equation}
\eta \propto \theta = 3 H. \label{etath}
\end{equation}
We will consider the case when
\begin{equation}
\eta = -\frac{3}{2 \kappa} H. \label{etath1}
\end{equation}
In this case from \eqref{Hn} and \eqref{Ven} one easily find
\begin{equation}
3 H^2 = \kappa \ve + C_4, \quad C_4 = {\rm const.} \label{shv}
\end{equation}
From \eqref{shv} it follows that at the initial state of
expansion, when $\ve$ is large, the Hubble constant is also large
and with the expansion of the Universe $H$ decreases as does
$\ve$. Inserting the relation \eqref{shv} into the Eqs. \eqref{Hn}
one finds
\begin{equation}
\int \frac{d H}{\sqrt{A H^2 + B H + C}} = t, \label{quad1}
\end{equation}
where, $A = - 1.5 (1 + \zeta)$,\, $B = 1.5 \kappa \xi$, and $C =
0.5 C_4 (\zeta - 1) - \Lambda$. For $\xi$ being a constant,
\eqref{quad1} admits sinusoidal solution, i.e., $H$ evolves
oscillatory. Further, from \eqref{tauH} one finds the expression
for $\tau$, which is exponential one accompanied by a sinusoidal
mode \cite{sahaprd1}.

               \section{Conclusion}
We investigated the cosmological solutions to the equations of
General Relativity for the homogeneous anisotropic Bianchi type I
model by taking into account dissipative processes due to
viscosity and Cosmological constant ($\Lambda$ term). A detailed
analysis showed that the viscosity, as well as the $\Lambda$ term
exhibit essential influence on the character of the solutions. The
classification of the solutions was pursued for the viscosity
being some power law of energy density, namely, $\eta = A
\ve^\alpha$ and $\xi = B \ve^\beta$. It was noticed that for
$\Lambda < 0$ the Universe expands forever with a logarithmic
velocity $H$, which, depending on the viscosity either becomes
constant or increases infinitely. In the process behavior of the
energy density $\ve$ is analogous to that of $H$ except the case
when $\ve \to 0$. For $\Lambda > 0$, beside the variants mentioned
above, there exists few other possibilities: contraction of the
Universe into a point, thus giving rise to a space-time
singularity; a regime of increasing oscillation corresponding to
suitable initial conditions. It was also noticed that a special
case with $\Lambda > 0$, \, $\eta = 0$ and $\xi H = {\rm const.}$
the model admits a singularity-free oscillatory mode of expansion.


\newcommand{\hnl}{\htmladdnormallink}


\vskip 5 cm


\myfigures{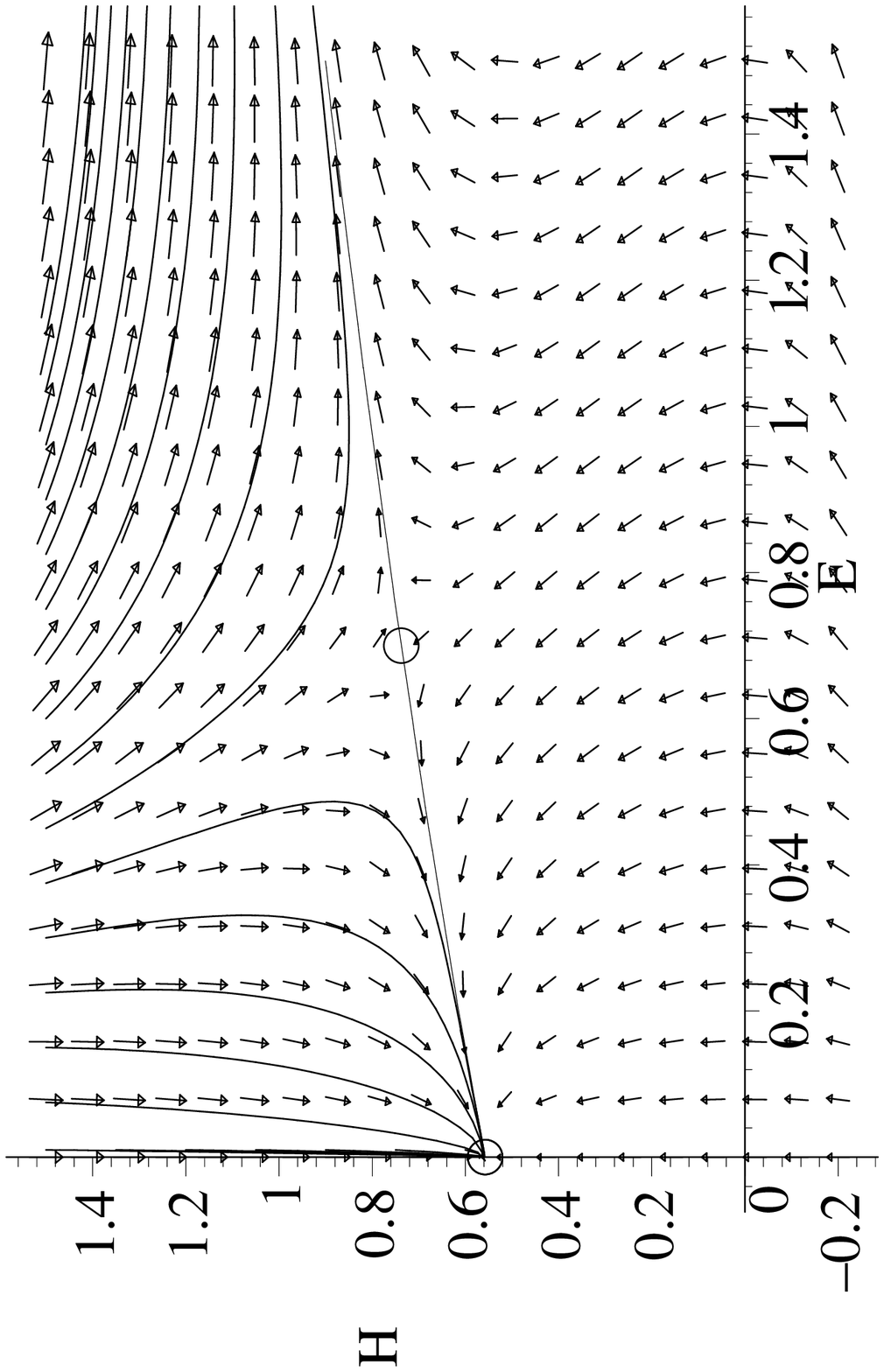}{0.30}{Phase diagram on $H-\ve$ plane for $\beta =
1.5$,\, $\Lambda = -.933$,\, $B = .720$.}{0.40}{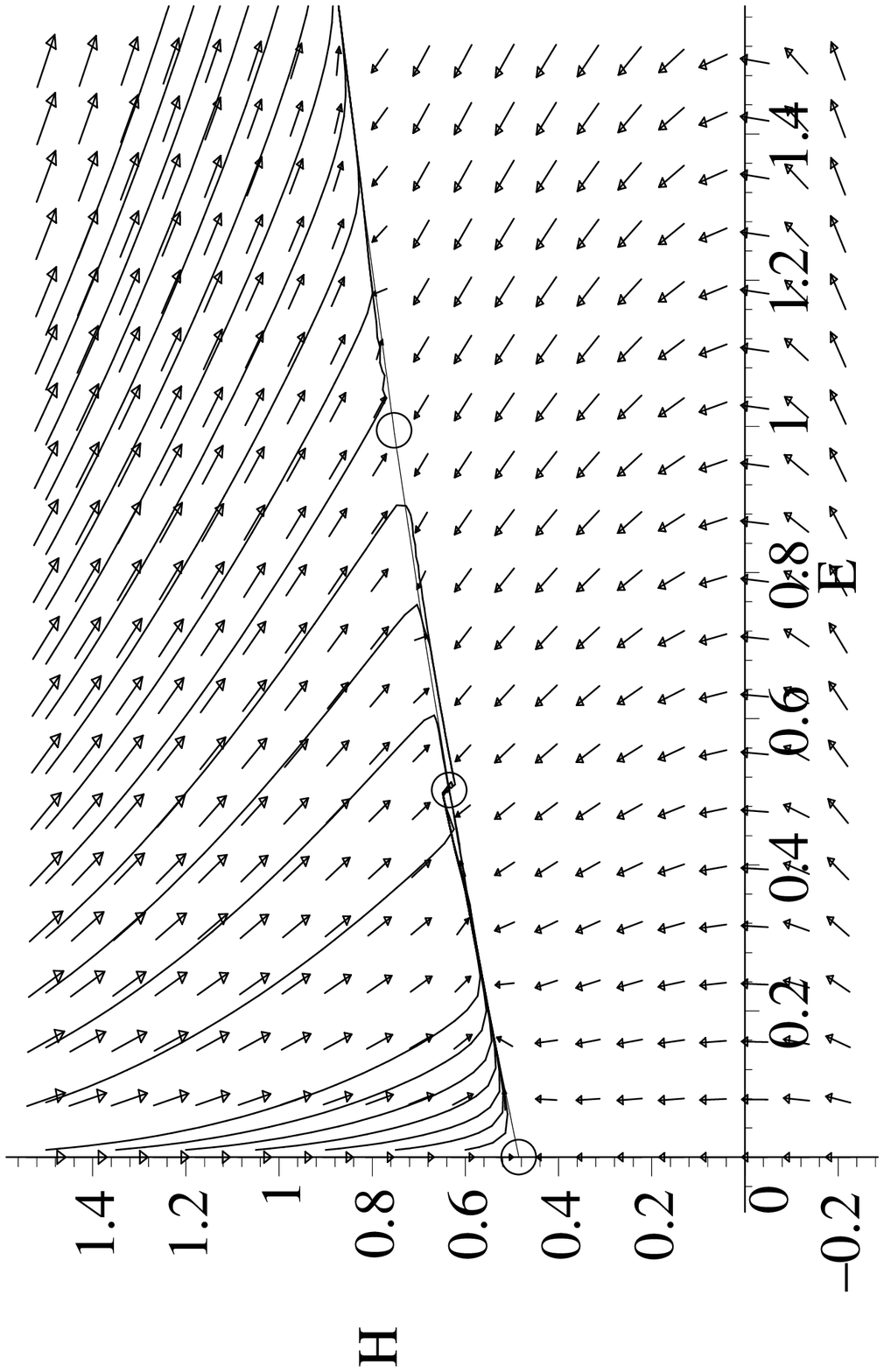}{0.30}{ Phase
diagram on $H-\ve$ plane for $\beta = .75$,\, $\Lambda = -.707$,\,
$B = .589$.}{0.40}

\myfigures{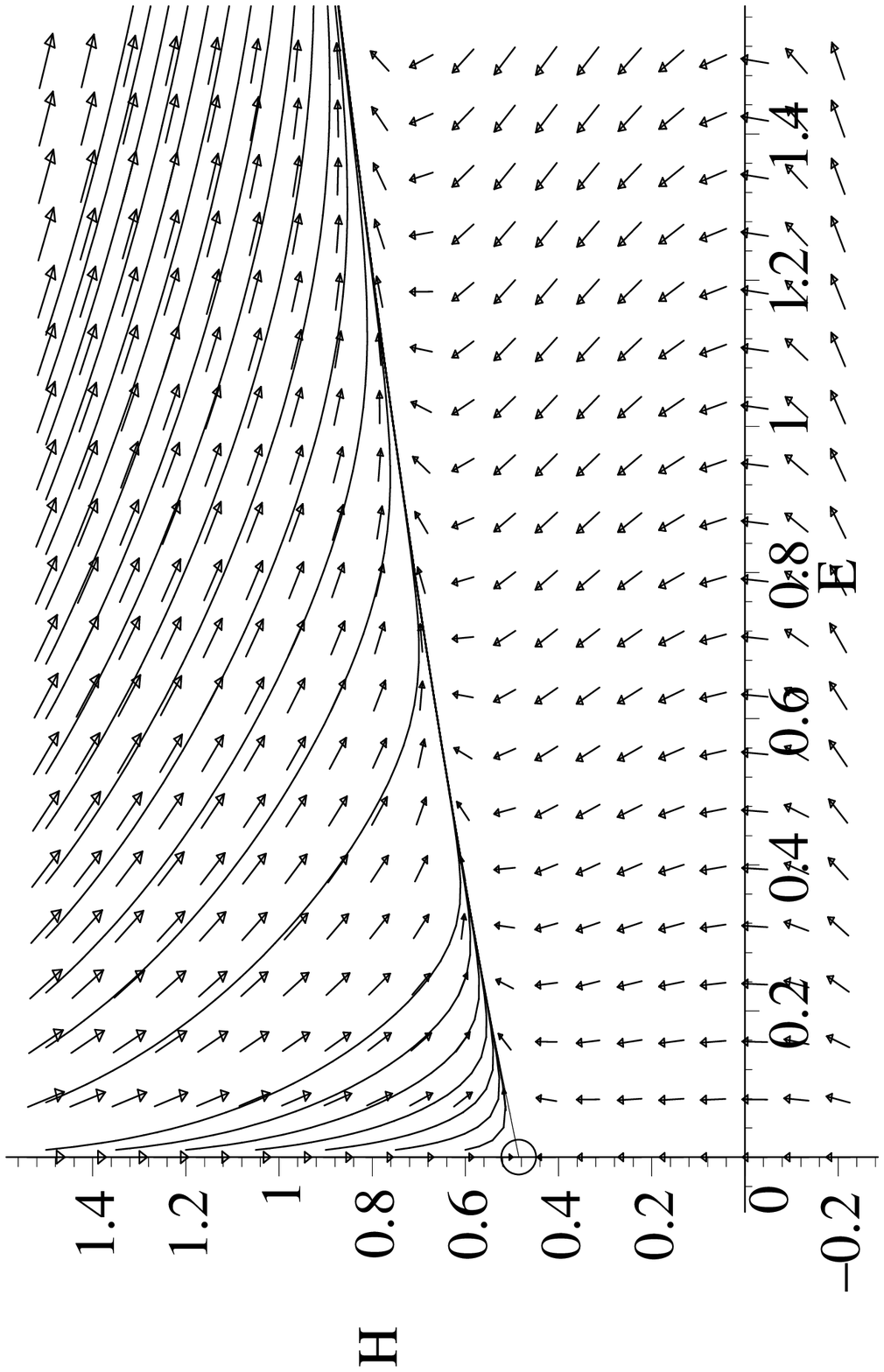}{0.30}{Phase diagram on $H-\ve$ plane for $\beta =
.75$,\, $\Lambda = -.707$,\, $B = .667$} {0.40}{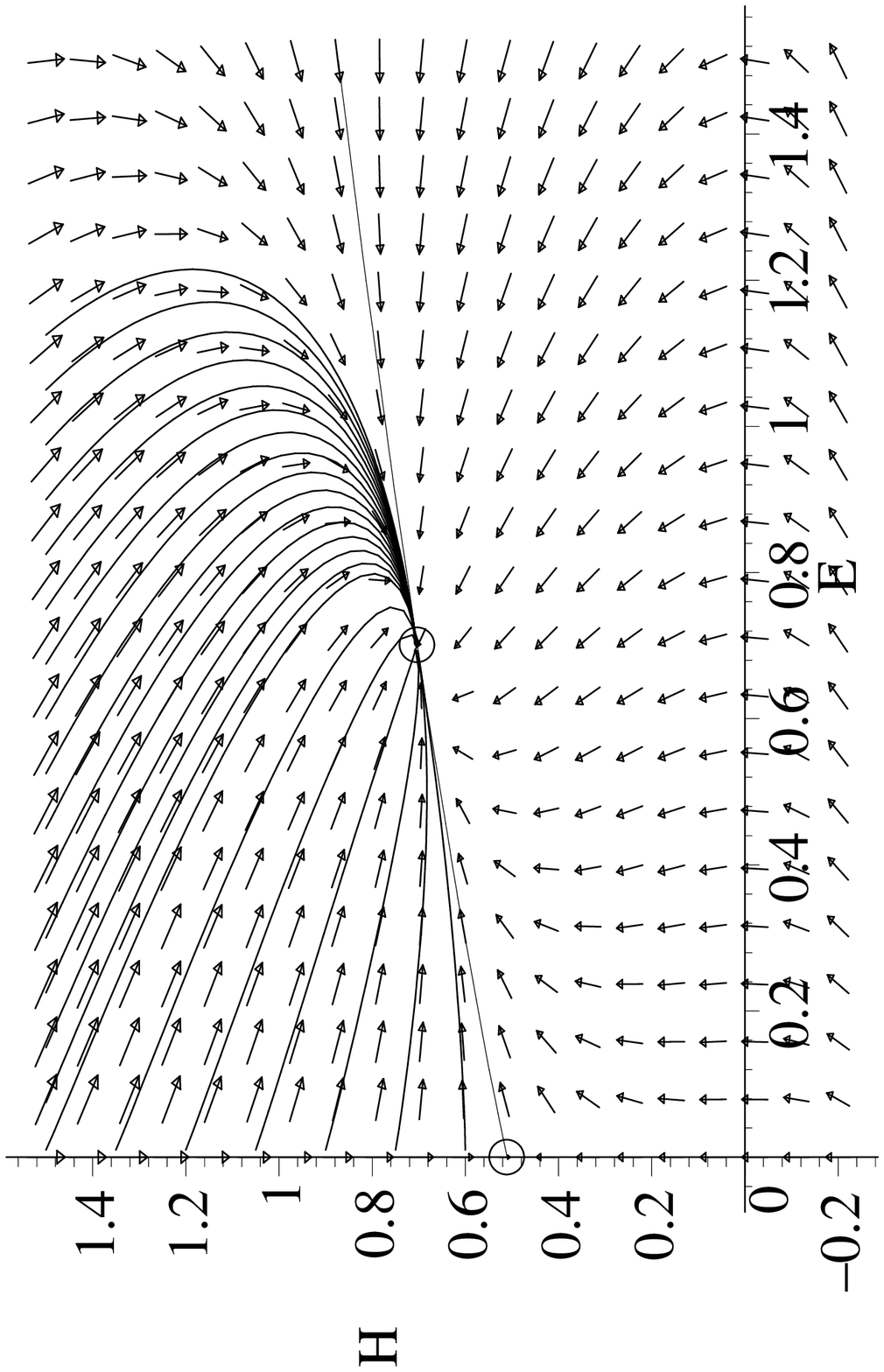}{0.30}{ Phase
diagram on $H-\ve$ plane for $\beta = .05$,\, $\Lambda = -.785$,\,
$B = .451$.}{0.40}

\myfigures{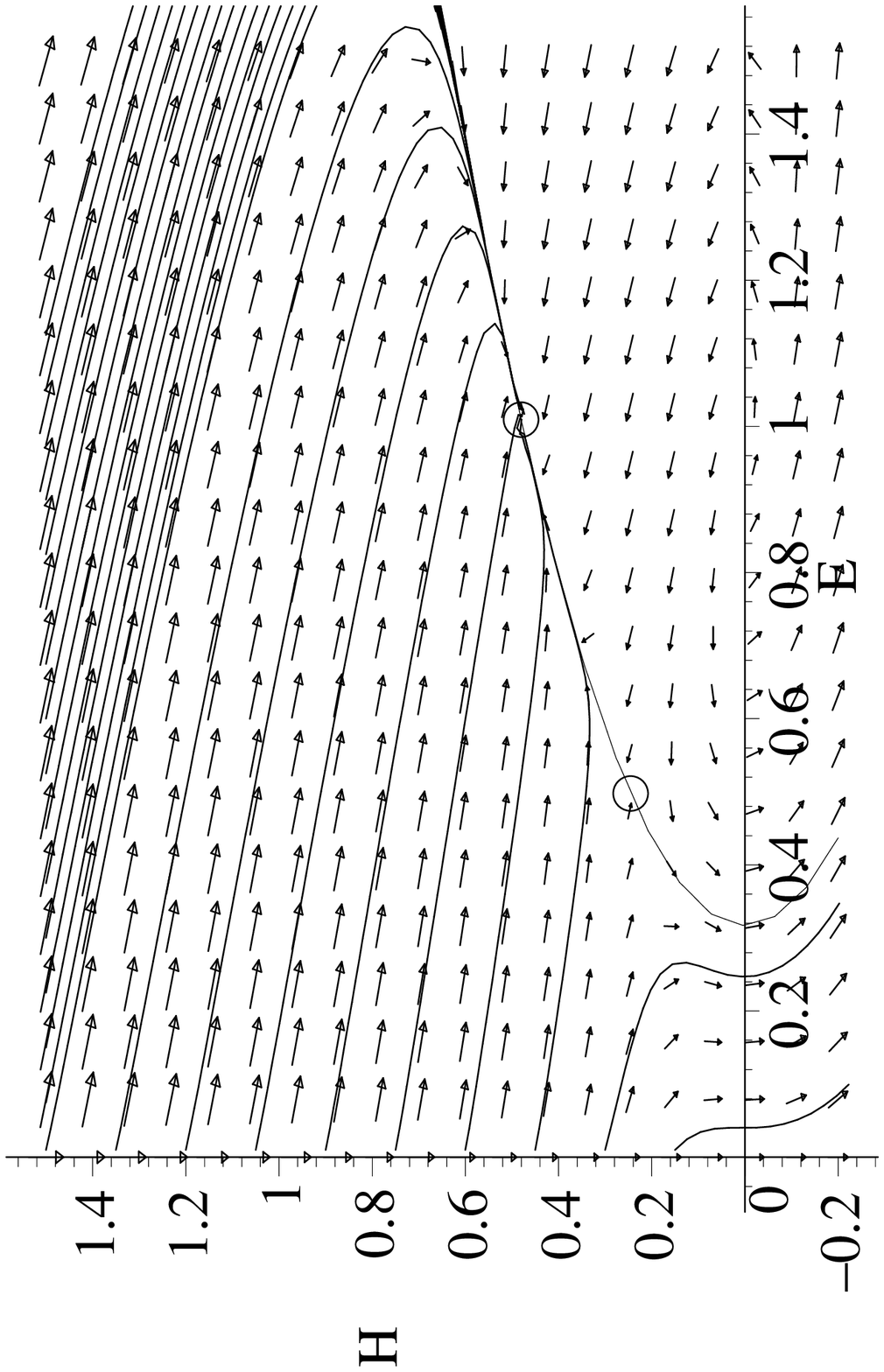}{0.30}{Phase diagram on $H-\ve$ plane for $\beta =
.05$,\, $\Lambda = .317$,\, $B = 0.933$} {0.40}{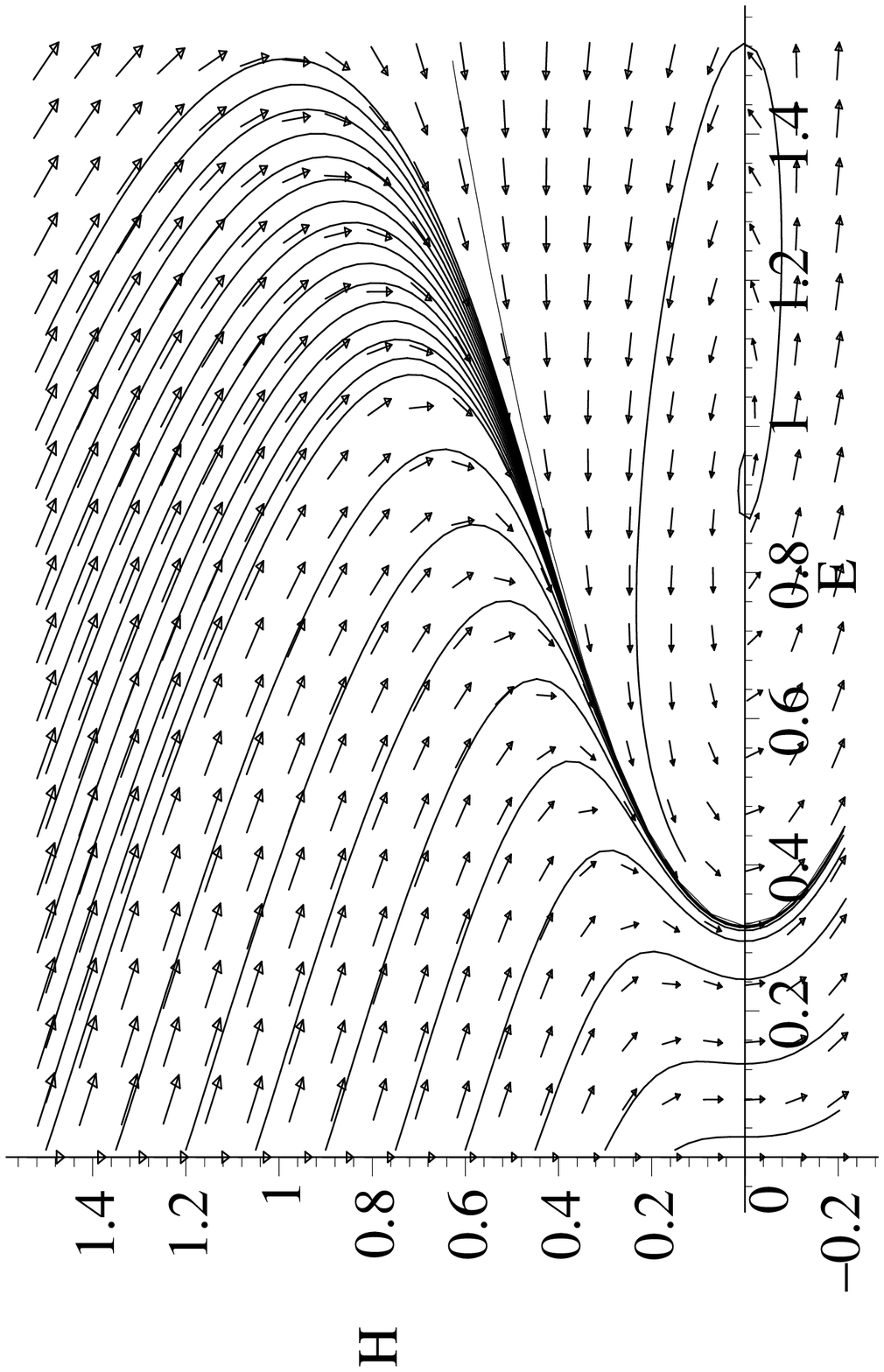}{0.30}{ Phase
diagram on $H-\ve$ plane for $\beta = .05$,\, $\Lambda = .317$,\,
$B = .667$.}{0.40}

\myfigures{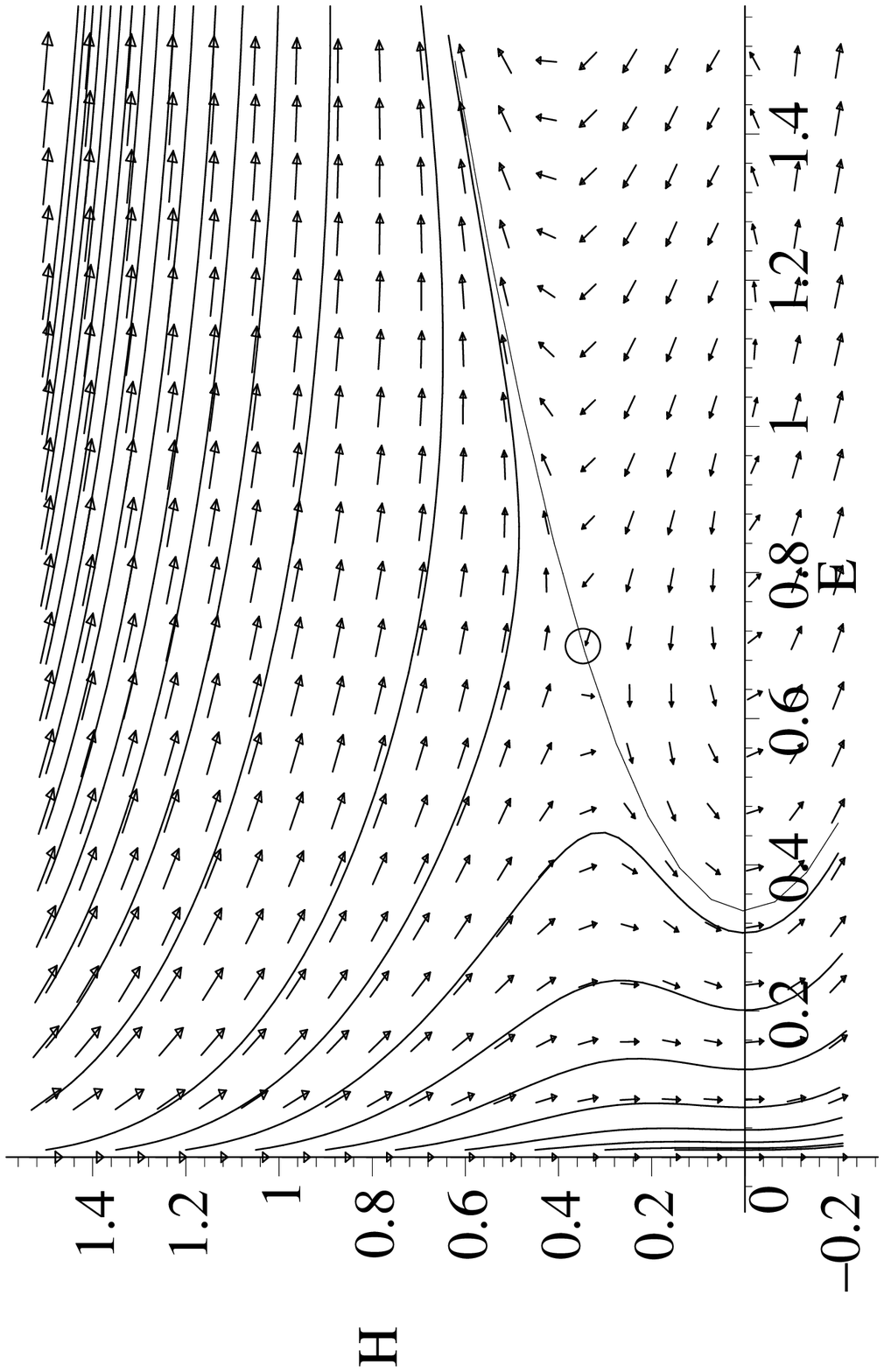}{0.30}{Phase diagram on $H-\ve$ plane for $\beta =
.75$,\, $\Lambda = .337$,\, $B = 1.169$.}{0.40}{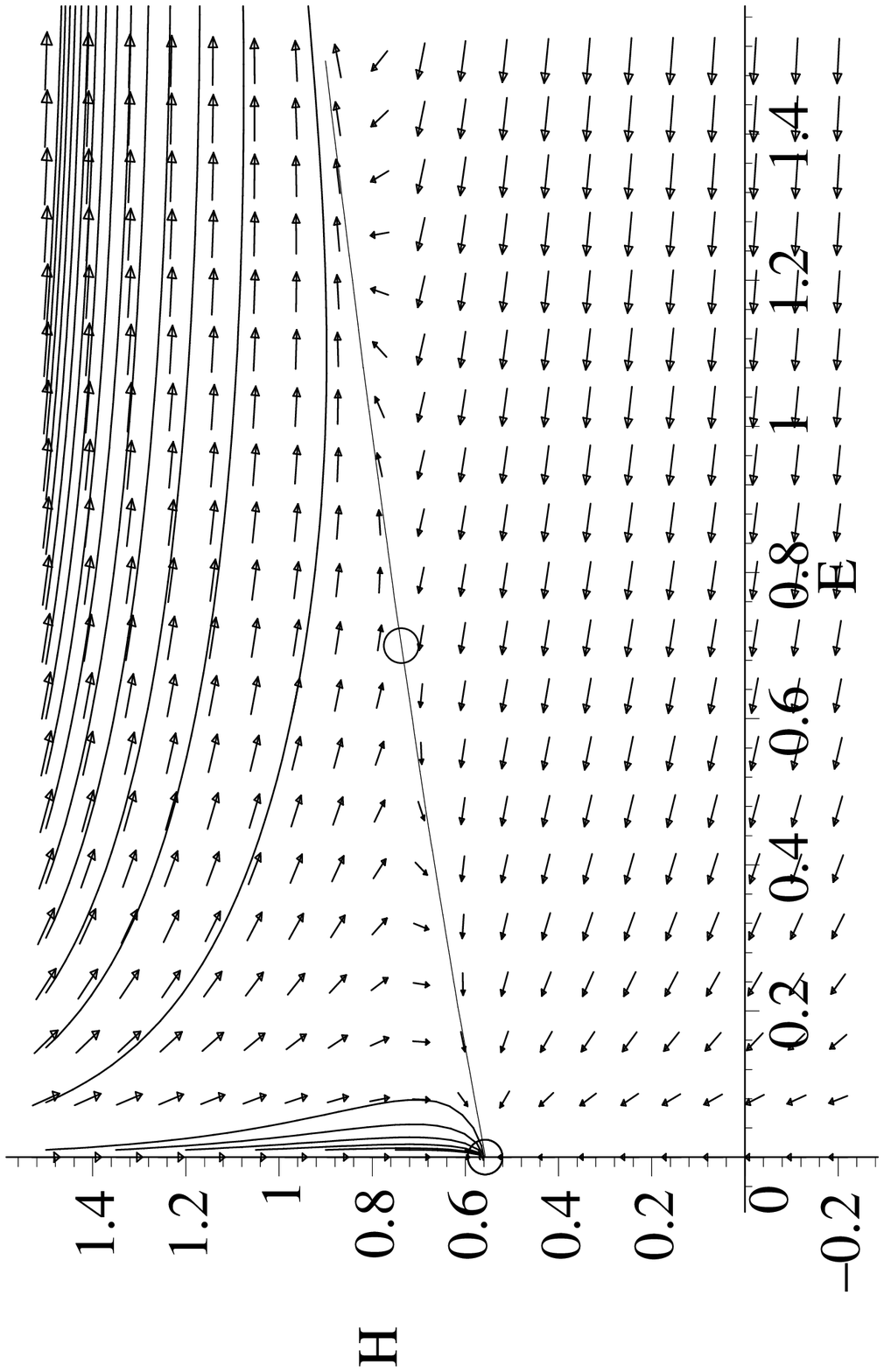}{0.30}{Phase
diagram on $H-\ve$ plane for $\beta = 1.5$,\, $\Lambda = -.933$,\,
$B = .720$,\,$A = 1,$\, $\alpha = 1$.}{0.40}


\myfigures{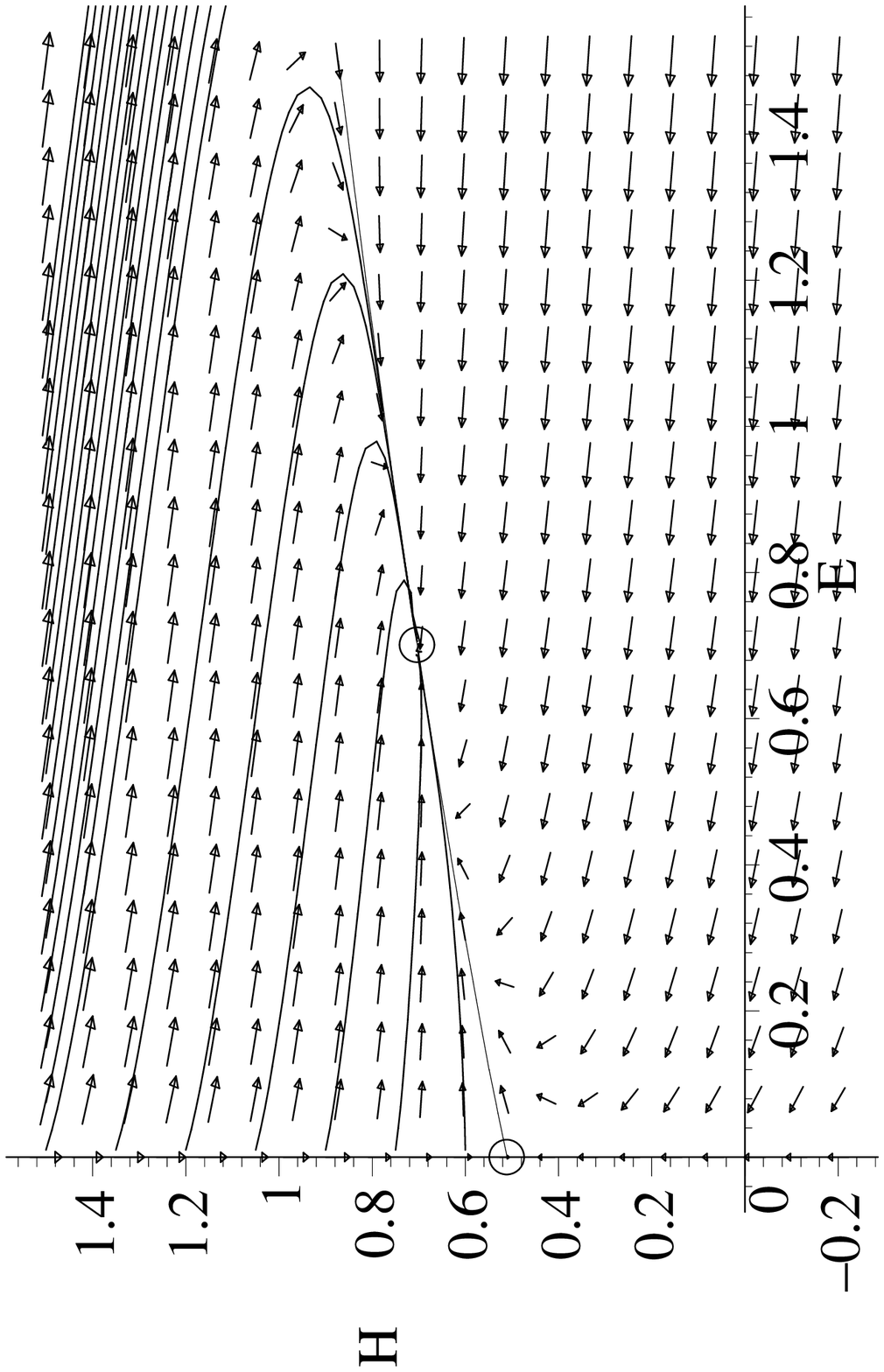}{0.30}{ Phase diagram on $H-\ve$ plane for $\beta =
.05$,\, $\Lambda = -.785$,\, $B = .451$,\,$A = 1,$\, $\alpha =
1$,\, $\kappa = 1.$}{0.40}{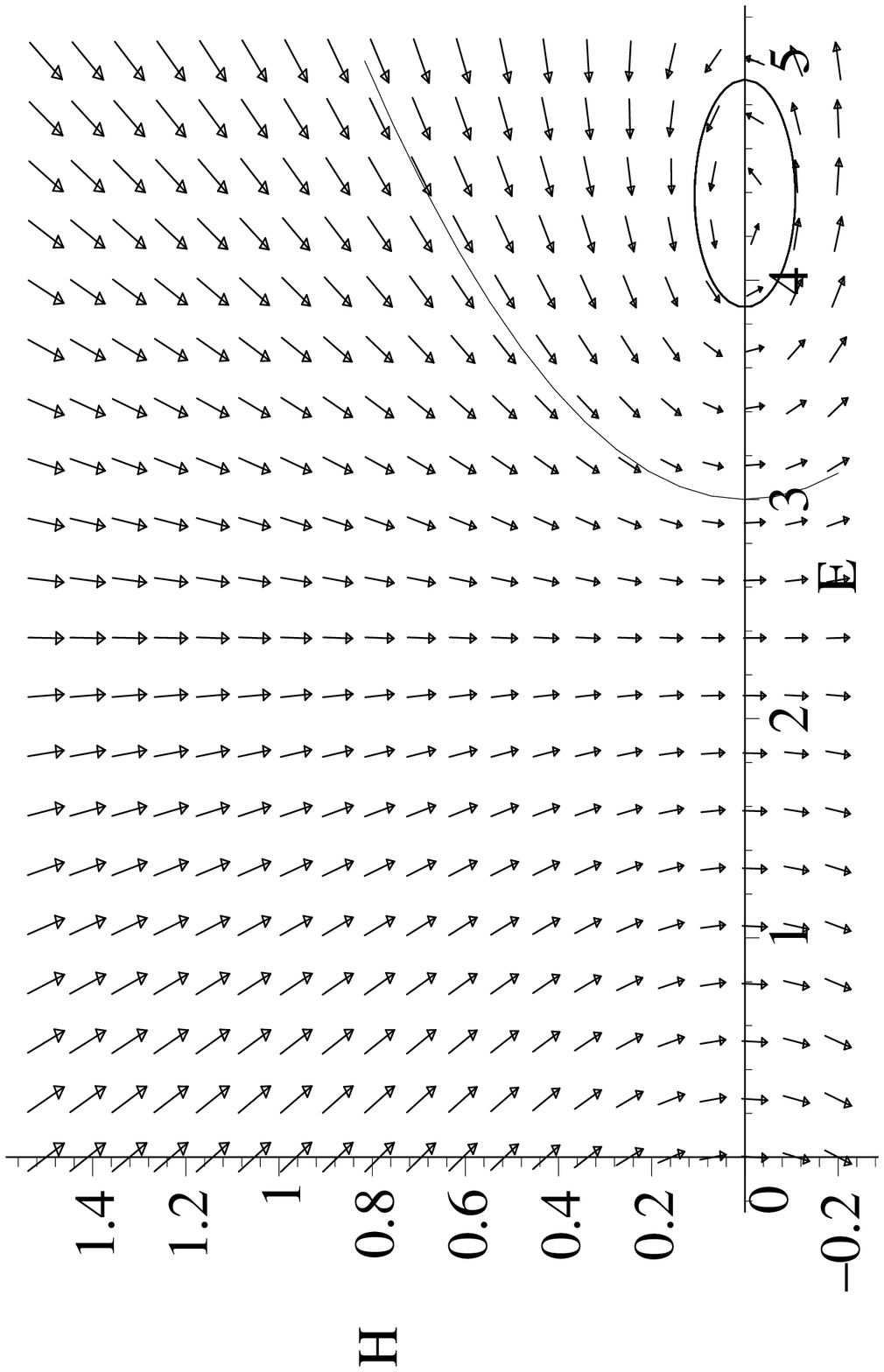}{0.30}{Phase diagram on $H-\ve$
plane for $\Lambda = 3$,\, $\zeta = 0.333$,\, $C_2 = 1,$\, $C_3 =
1.$}{0.40}

\myfigure{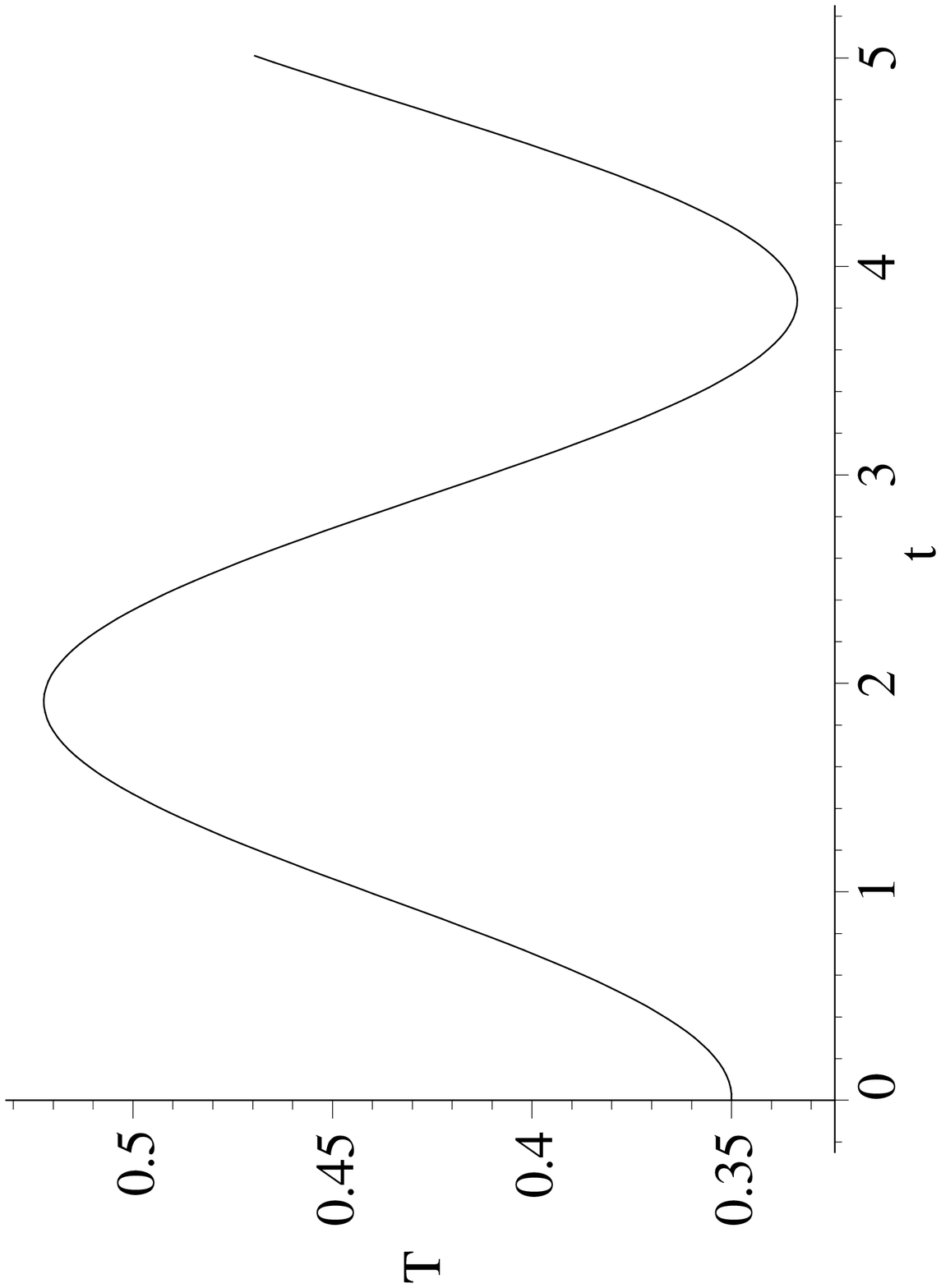}{0.30}{Evolution of the BI Universe corresponding to
the phase diagram given in Fig. \ref{he10}. As one sees, the BI
Universe in this case undergoes an oscillatory mode of
expansion}{0.70}

\myf{0.20}{0.27} {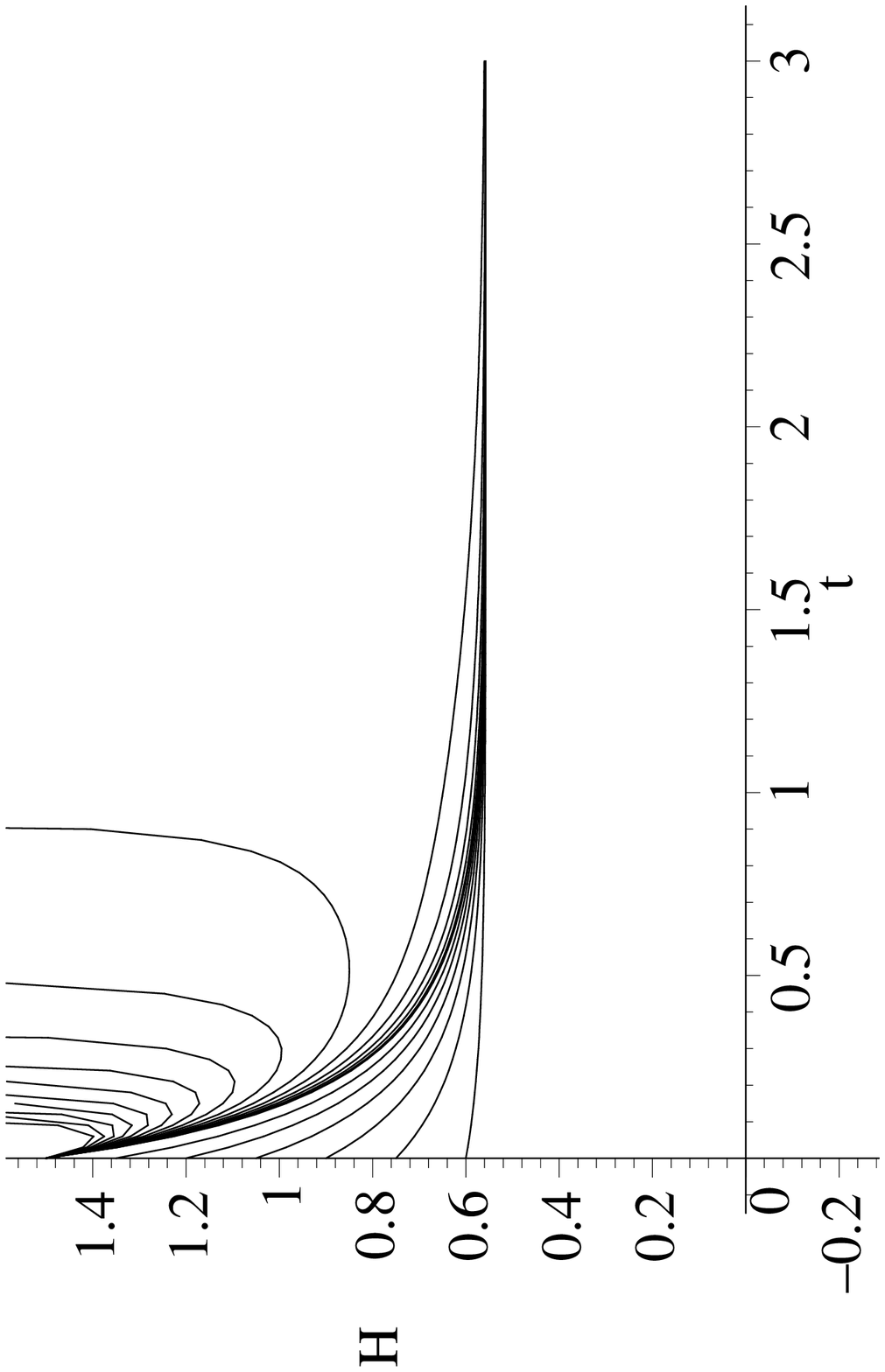}{Evolution of the Hubble constant $H$ with
parameters as in Fig. \ref{he1}.} {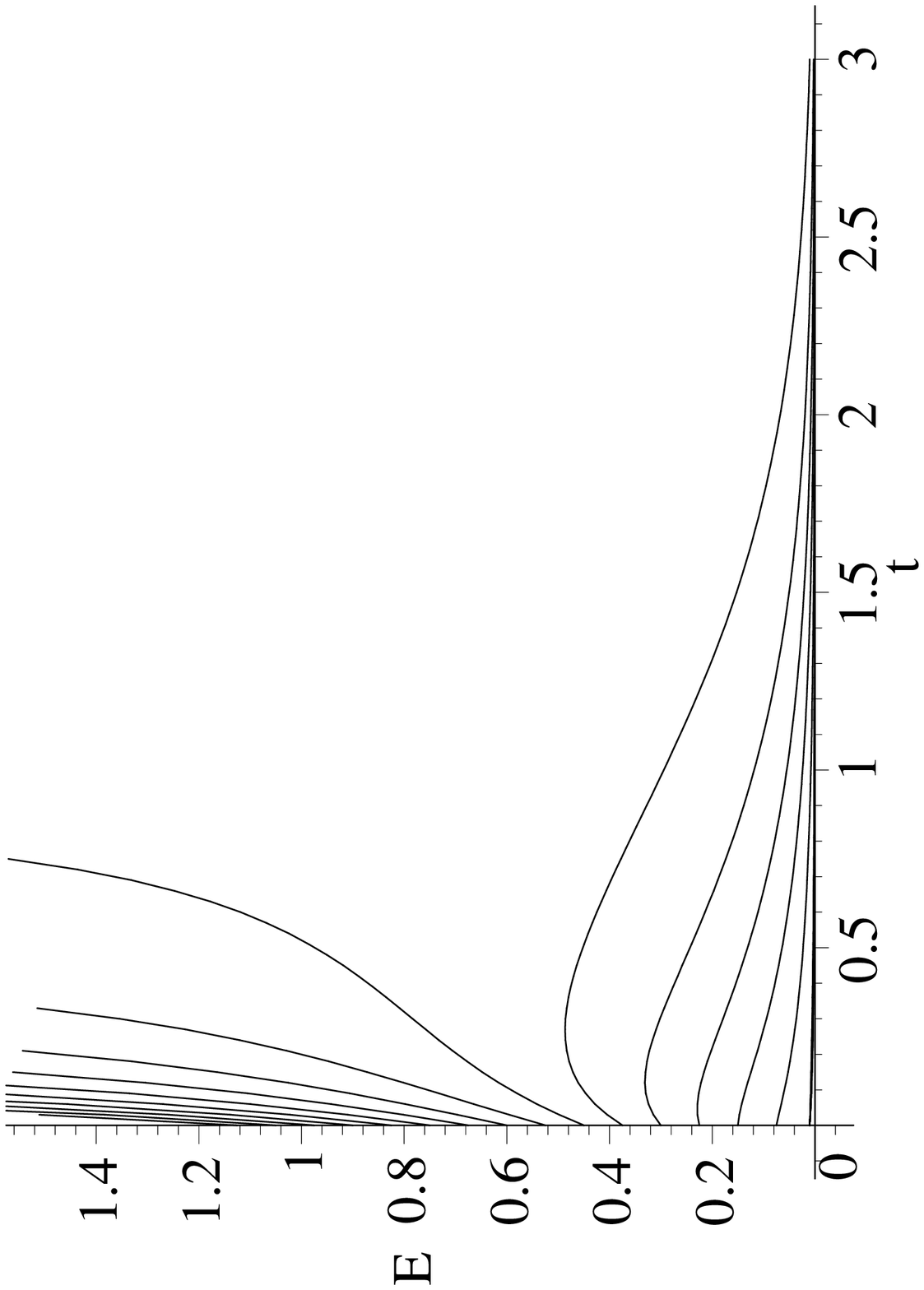}{Evolution of the energy
density $\ve$ with parameters as in Fig. \ref{he1}.}{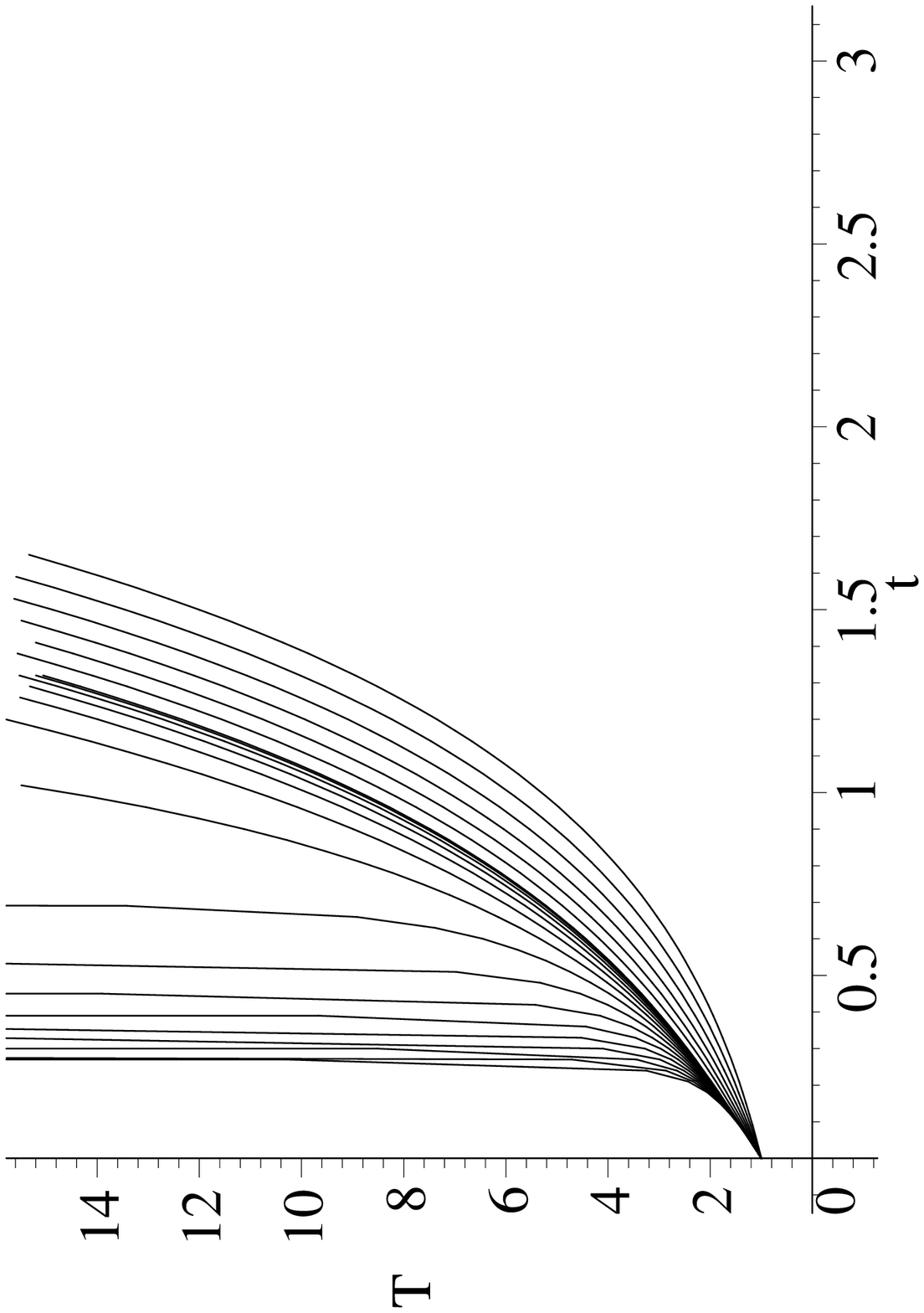}{Evolution
of the volume scale $\tau$ with parameters as in Fig. \ref{he1}.}

\myf{0.20}{0.27}{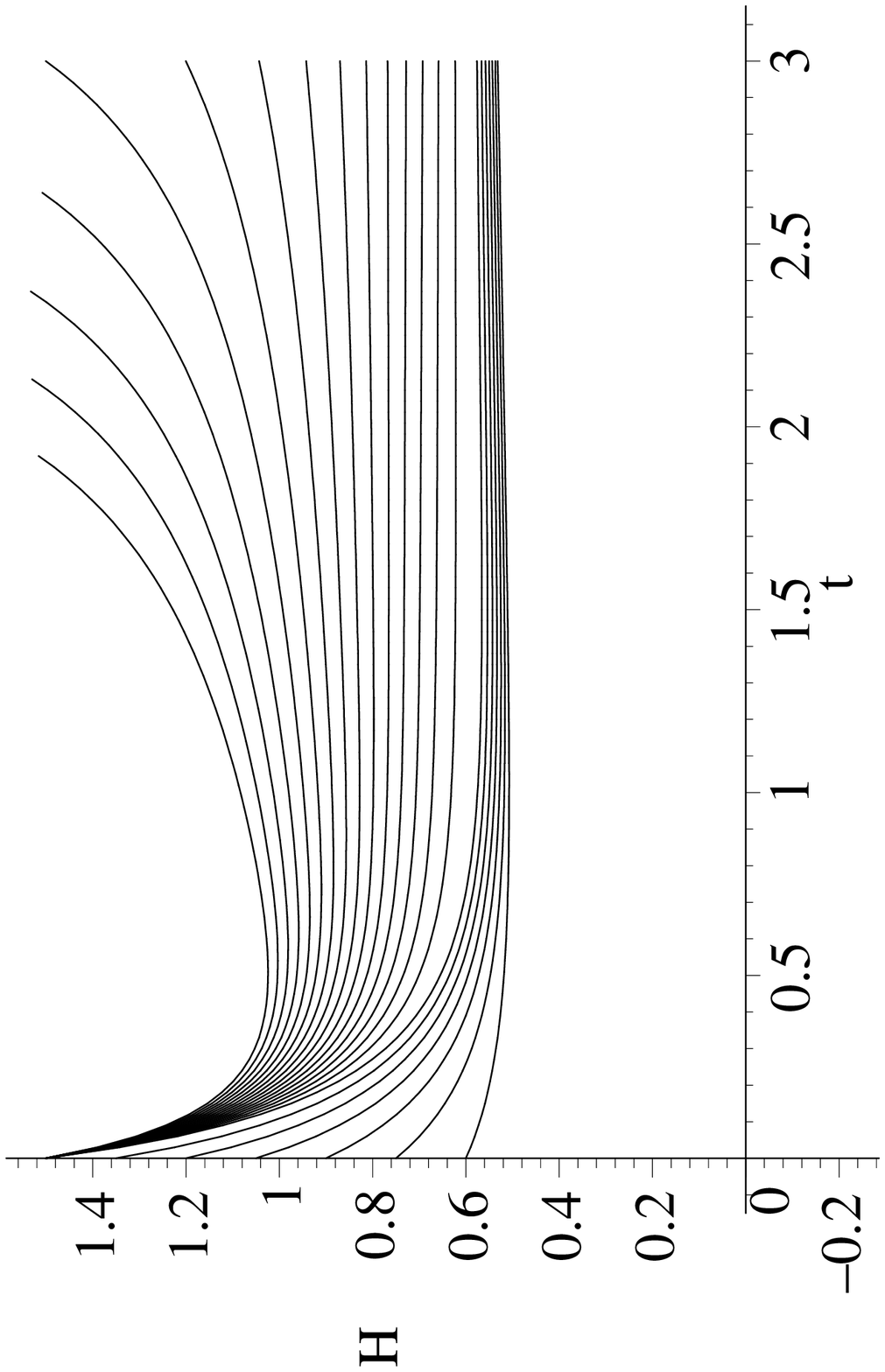}{Evolution of the Hubble constant $H$ with
parameters as in Fig. \ref{he2}.} {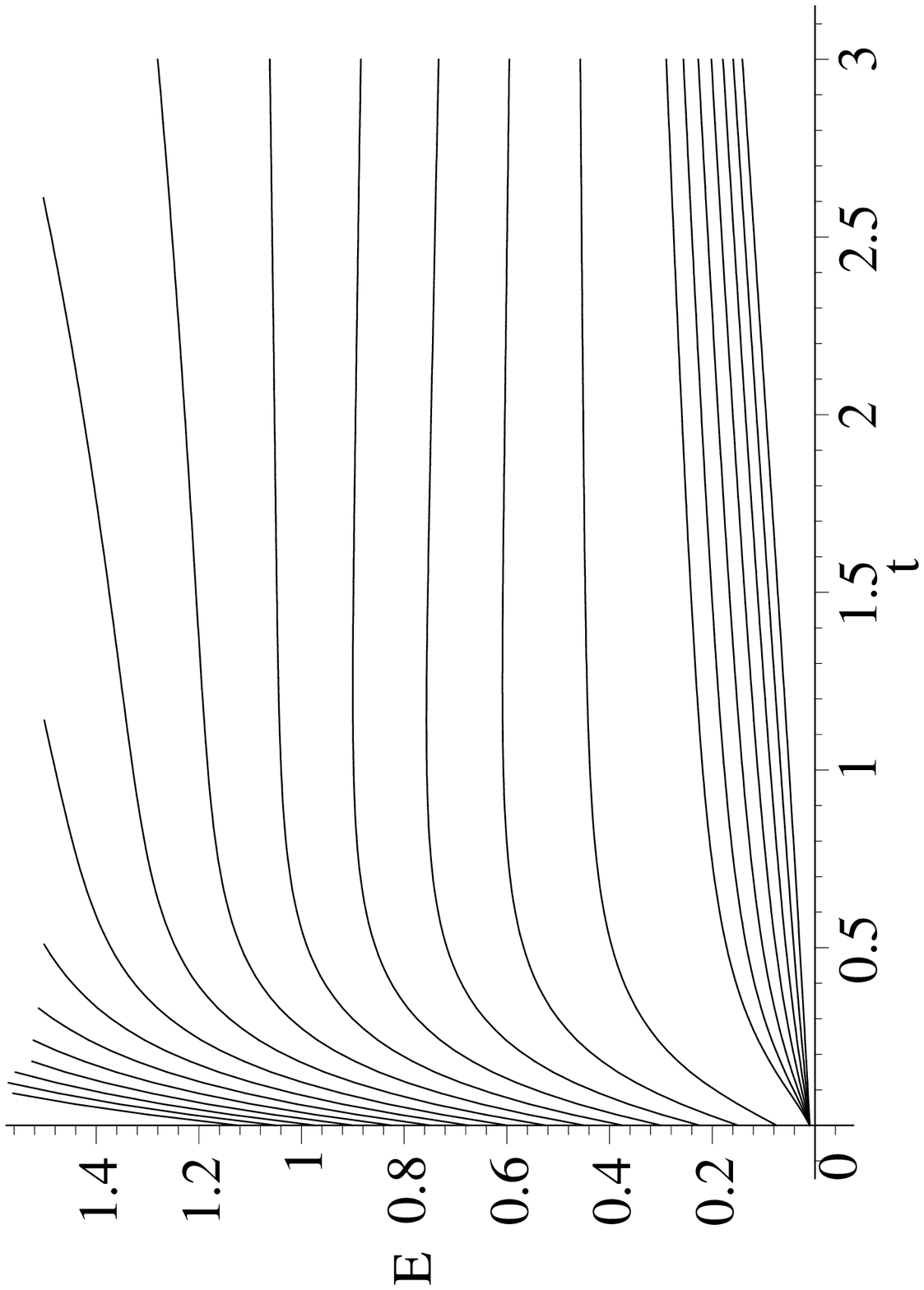}{Evolution of the energy
density $\ve$ with parameters as in Fig. \ref{he2}.}{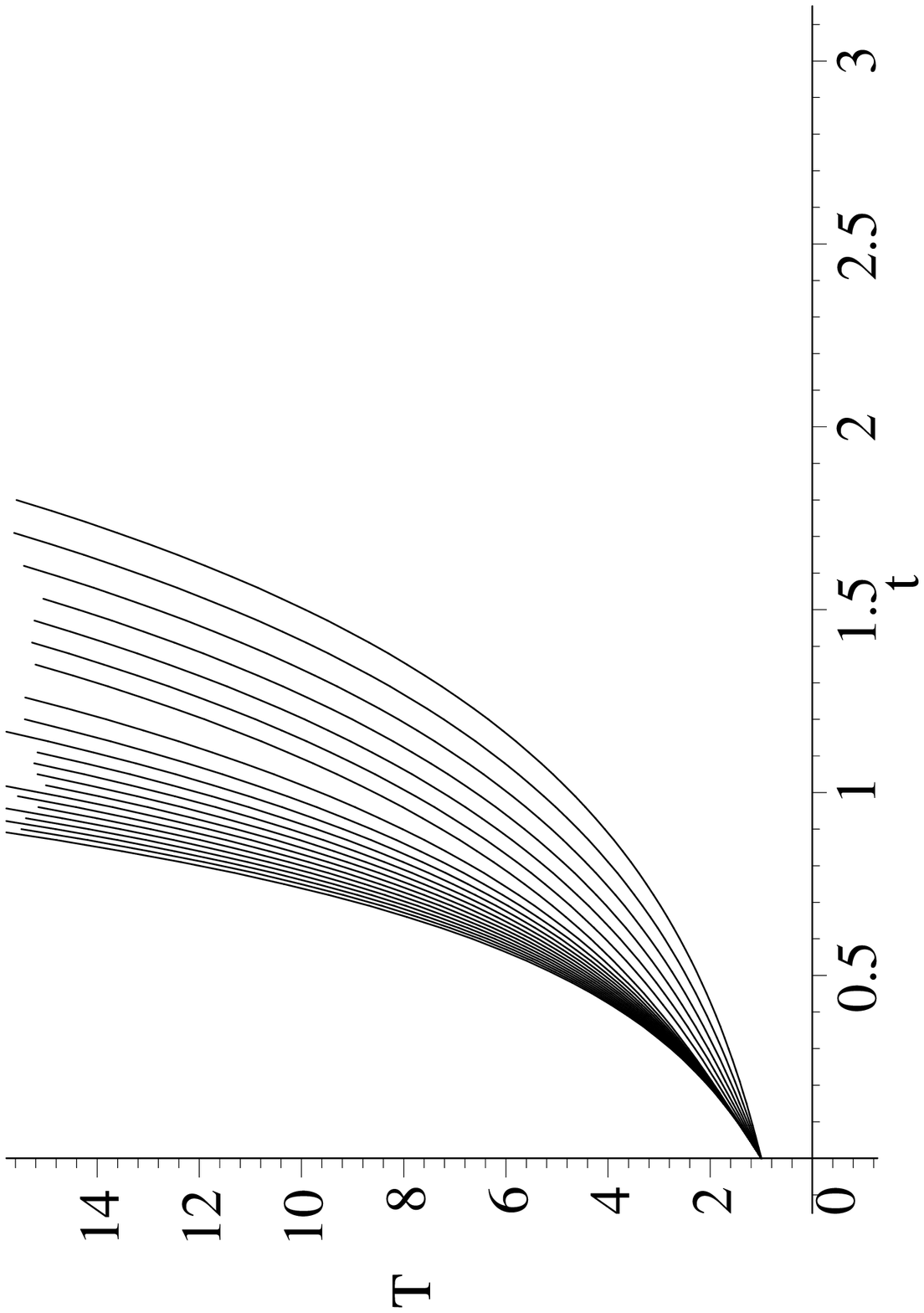}{Evolution
of the volume scale $\tau$ with parameters as in Fig. \ref{he2}.}

\myf{0.20}{0.27}{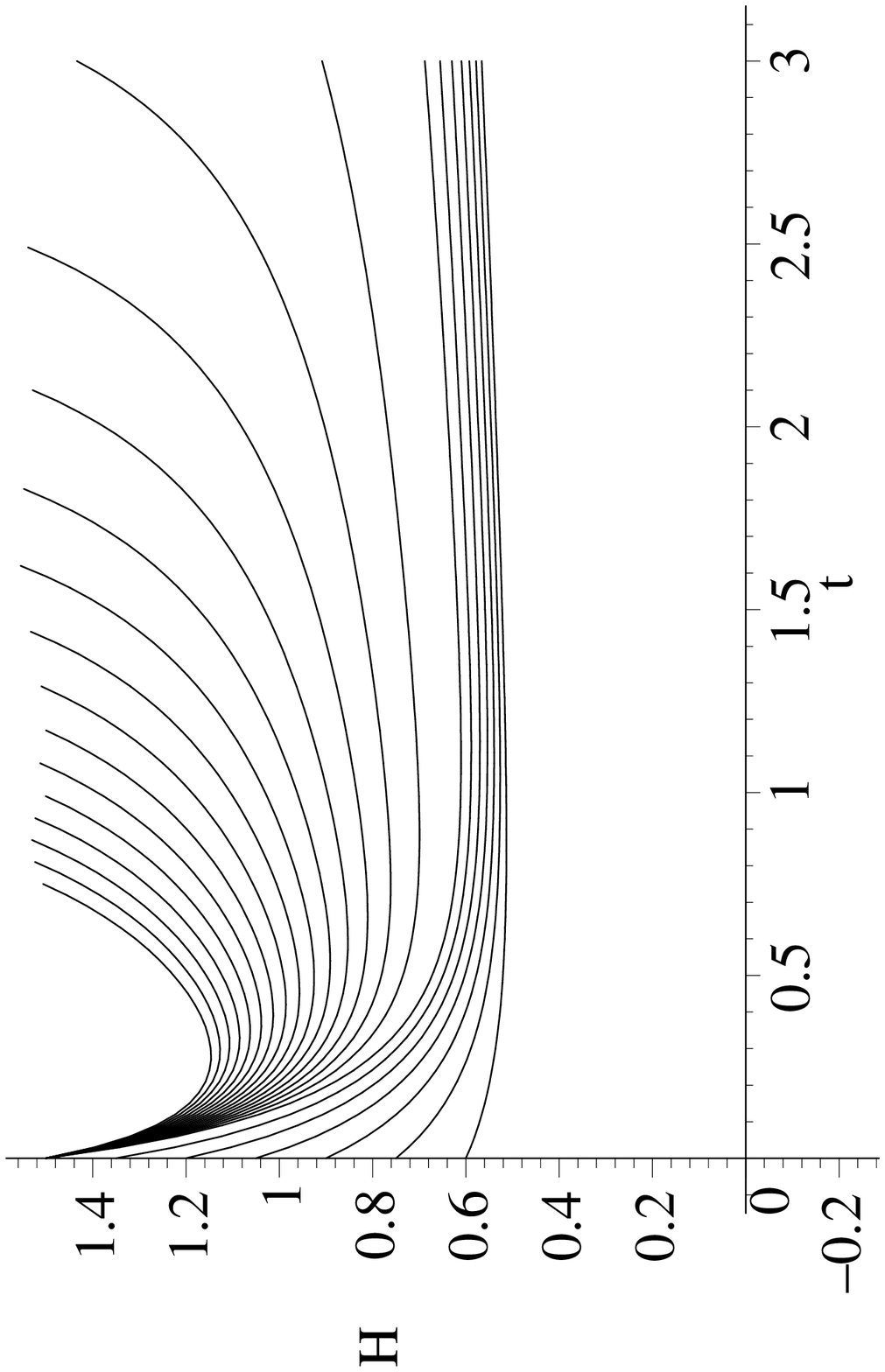}{Evolution of the Hubble constant $H$ with
parameters as in Fig. \ref{he3}.}{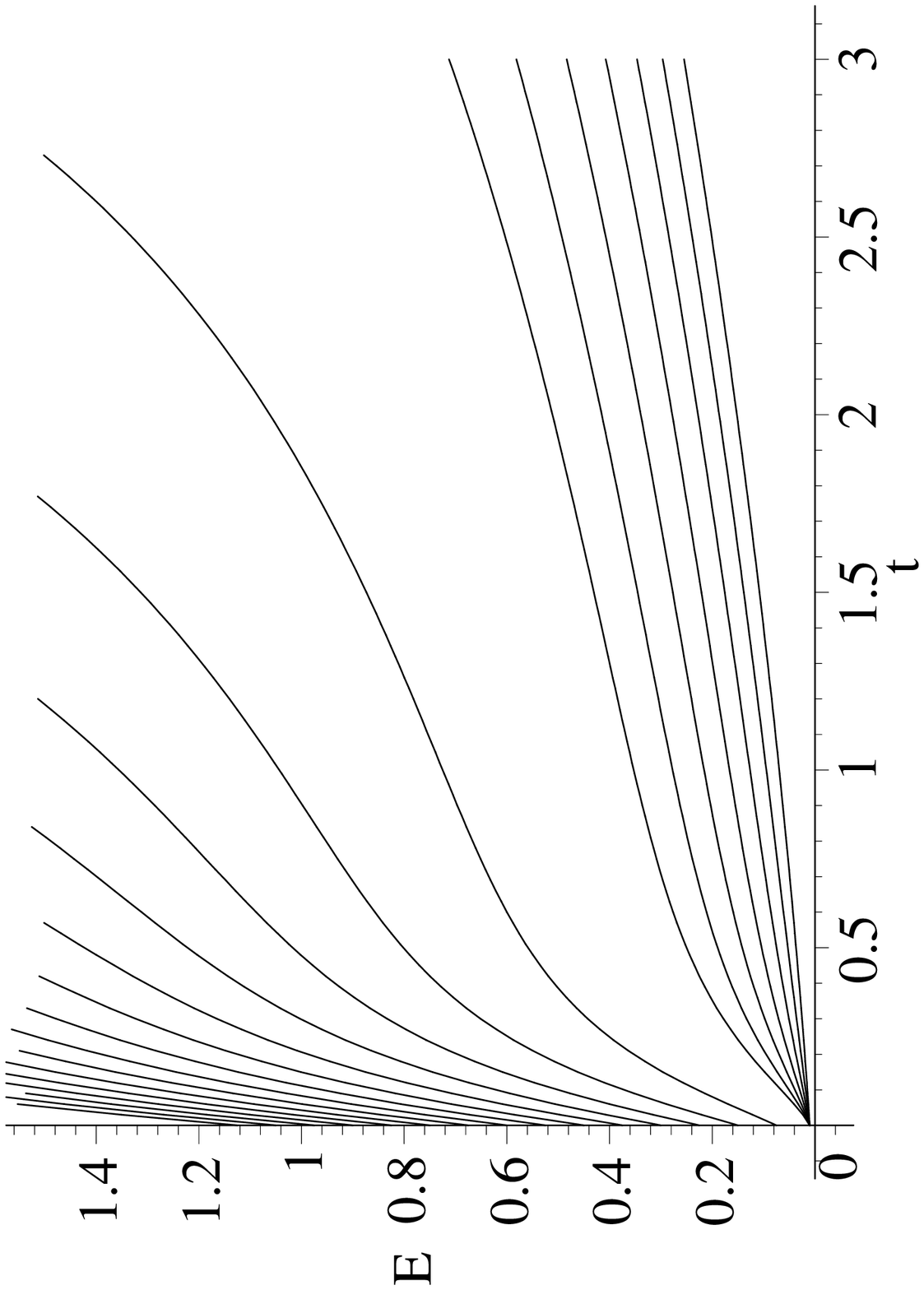}{Evolution of the energy
density $\ve$ with parameters as in Fig. \ref{he3}.}{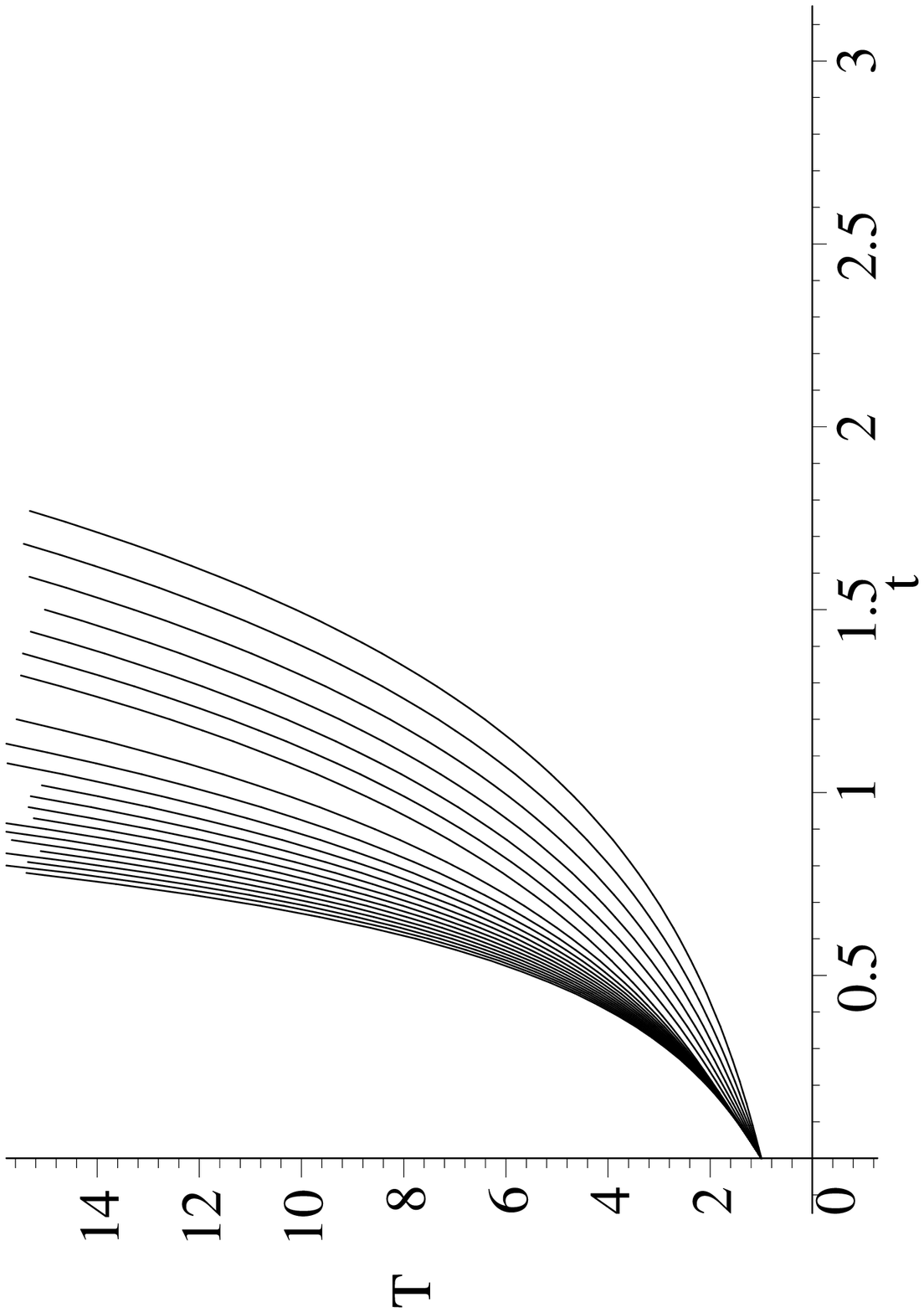}{Evolution
of the volume scale $\tau$ with parameters as in Fig. \ref{he3}.}

\myf{0.20}{0.27}{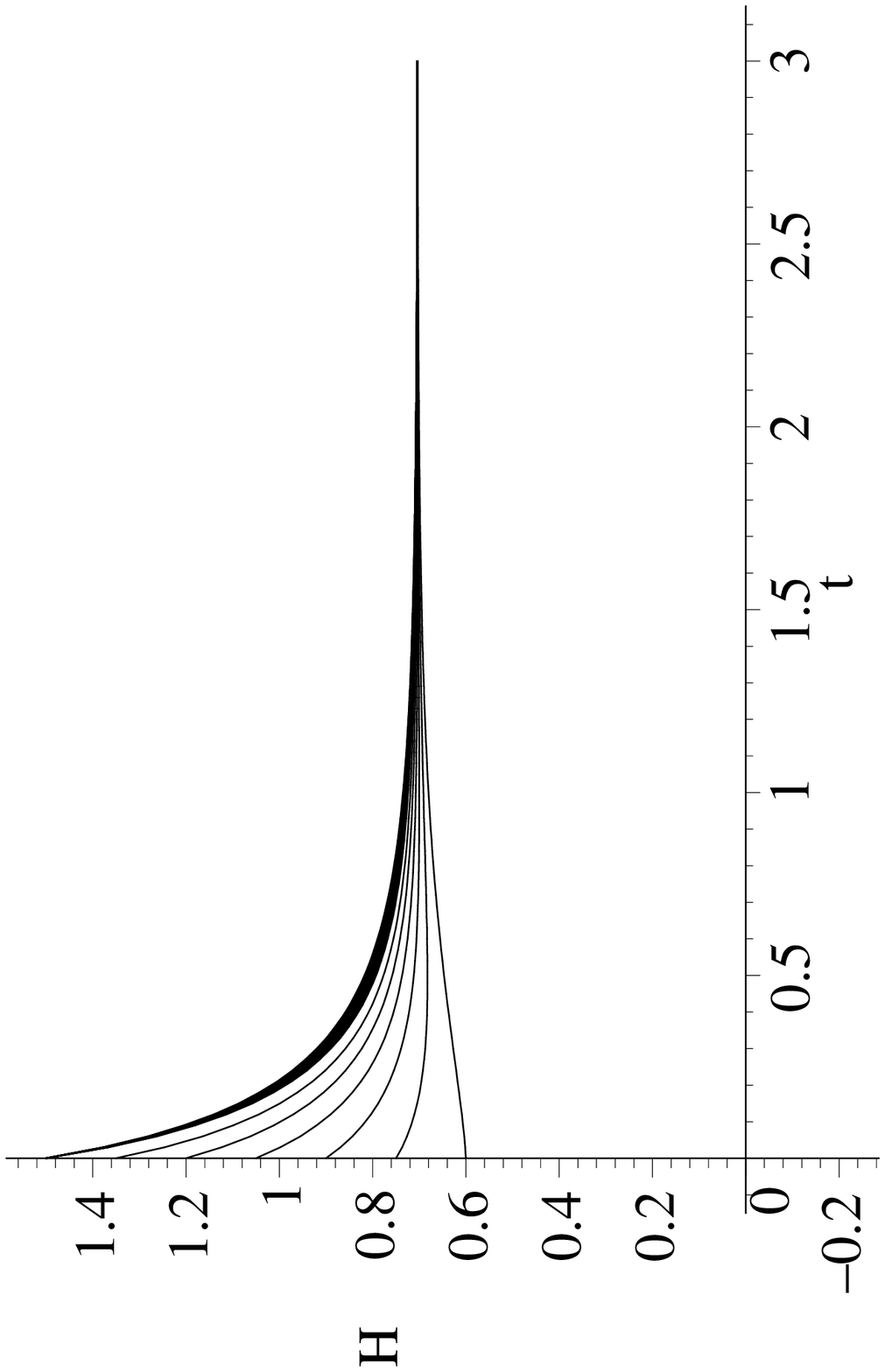}{Evolution of the Hubble constant $H$ with
parameters as in Fig. \ref{he4}.}{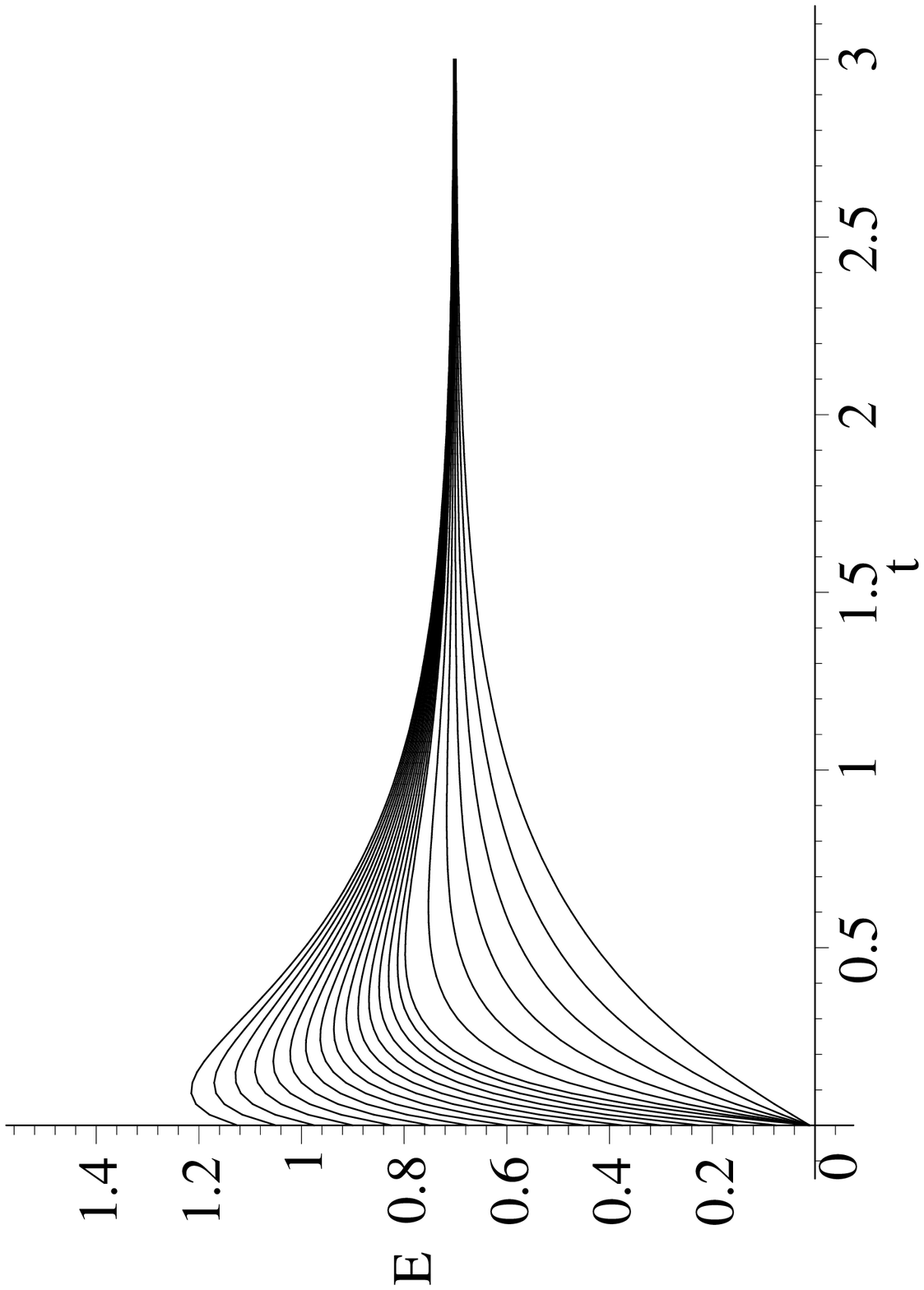}{Evolution of the energy
density $\ve$ with parameters as in Fig. \ref{he4}.}{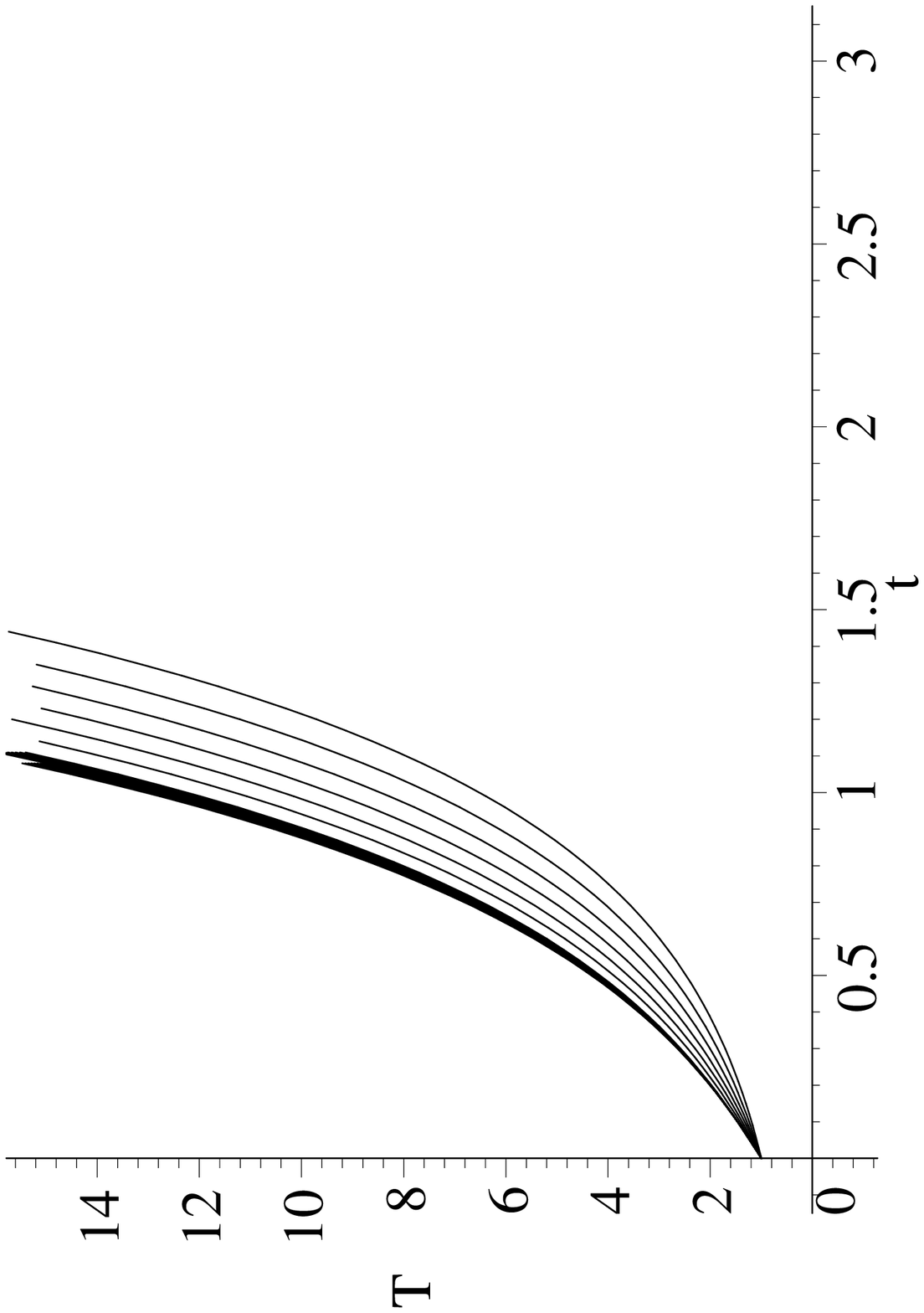}{Evolution
of the volume scale $\tau$ with parameters as in Fig. \ref{he4}.}

\myf{0.20}{0.27}{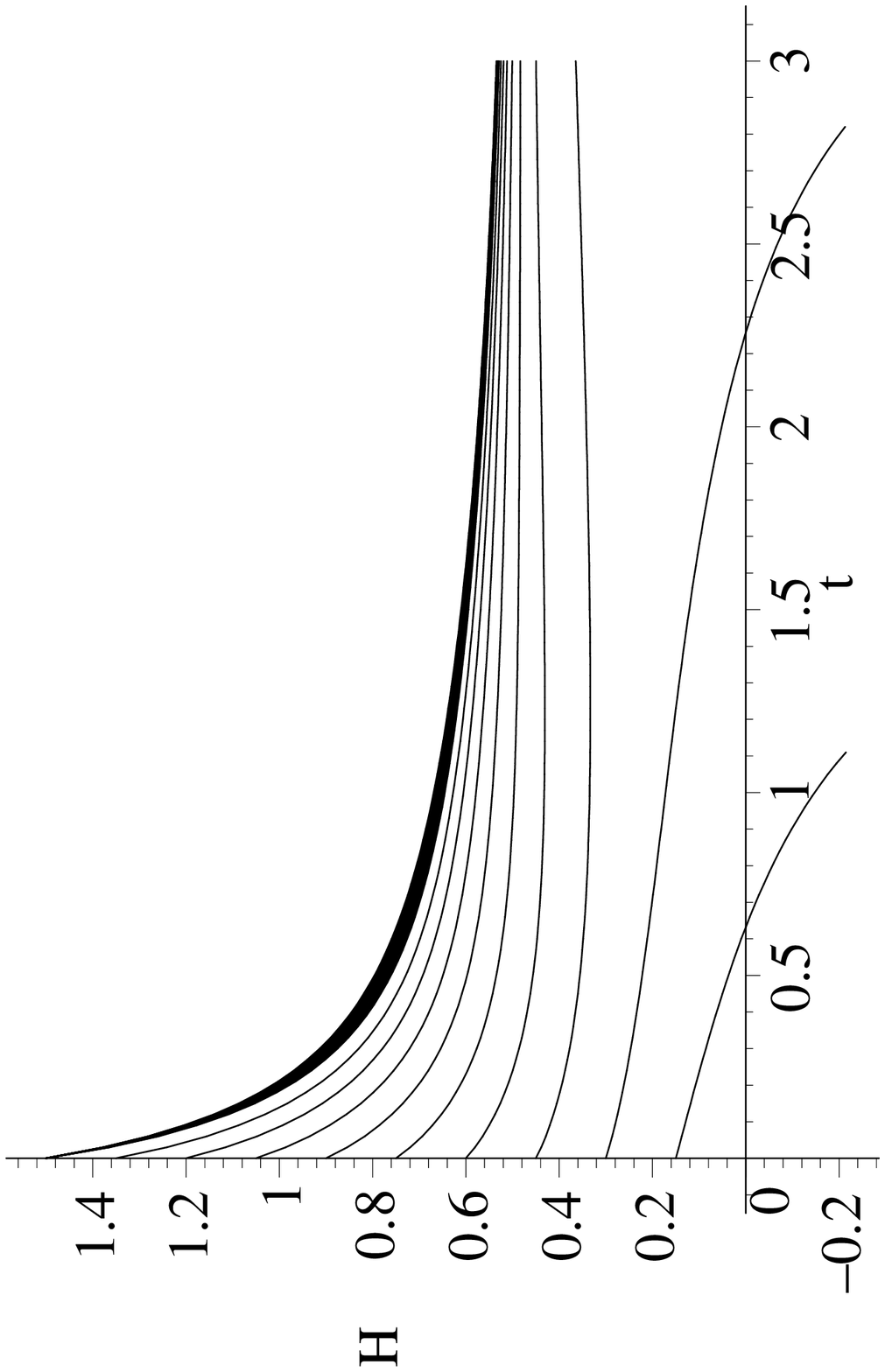}{Evolution of the Hubble constant $H$ with
parameters as in Fig. \ref{he5}.}{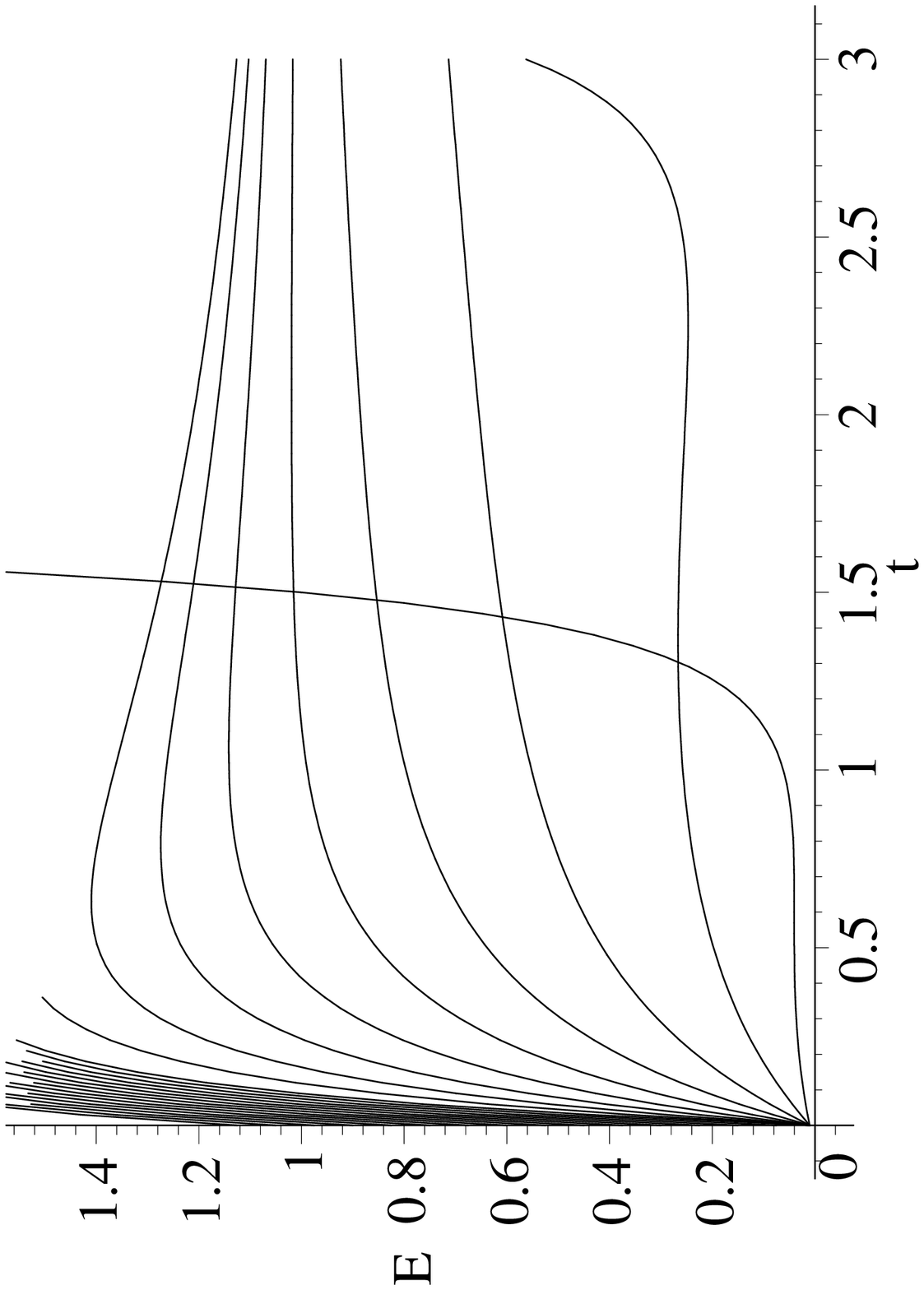}{Evolution of the energy
density $\ve$ with parameters as in Fig. \ref{he5}.}{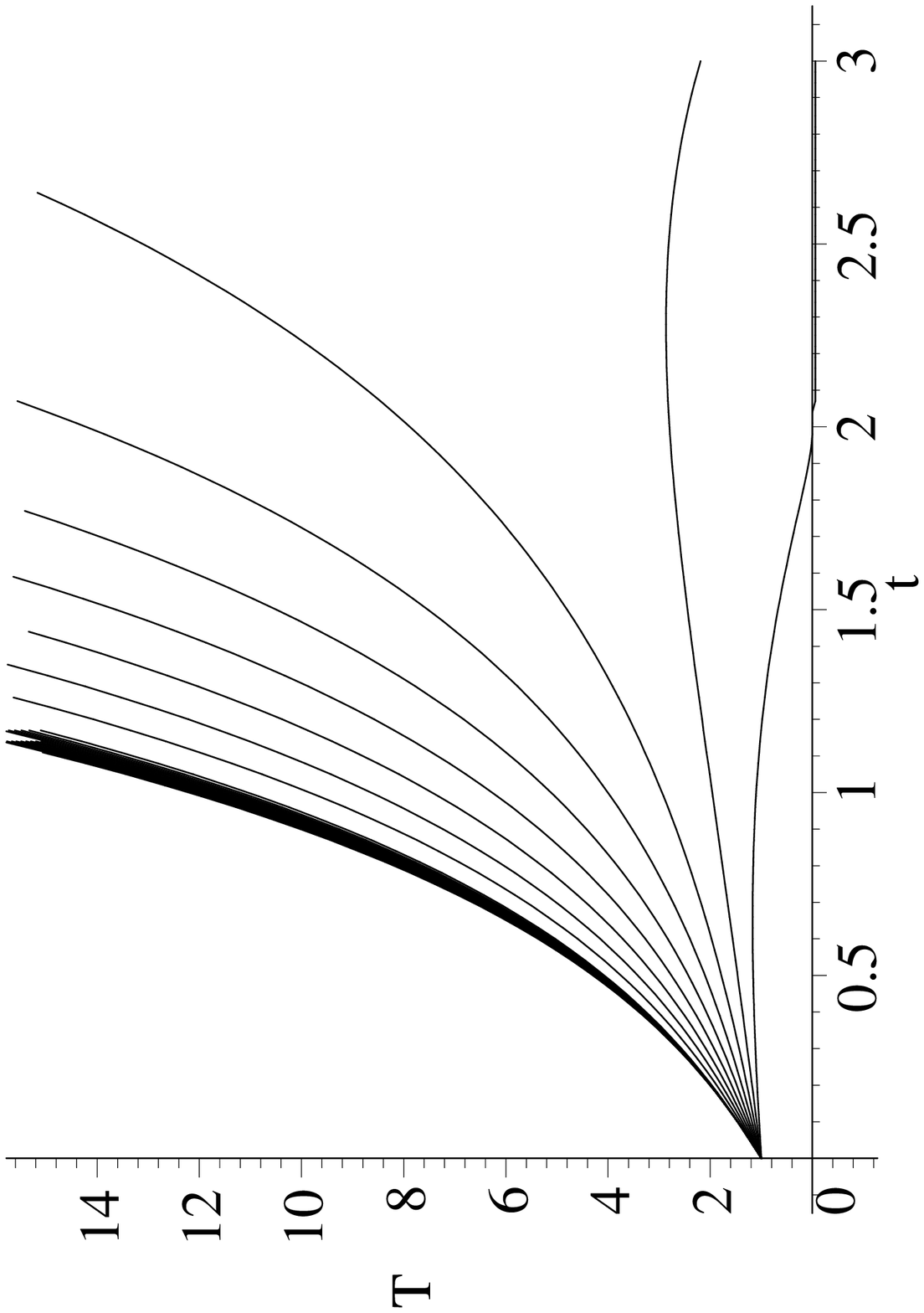}{Evolution
of the volume scale $\tau$ with parameters as in Fig. \ref{he5}.}

\myf{0.20}{0.27}{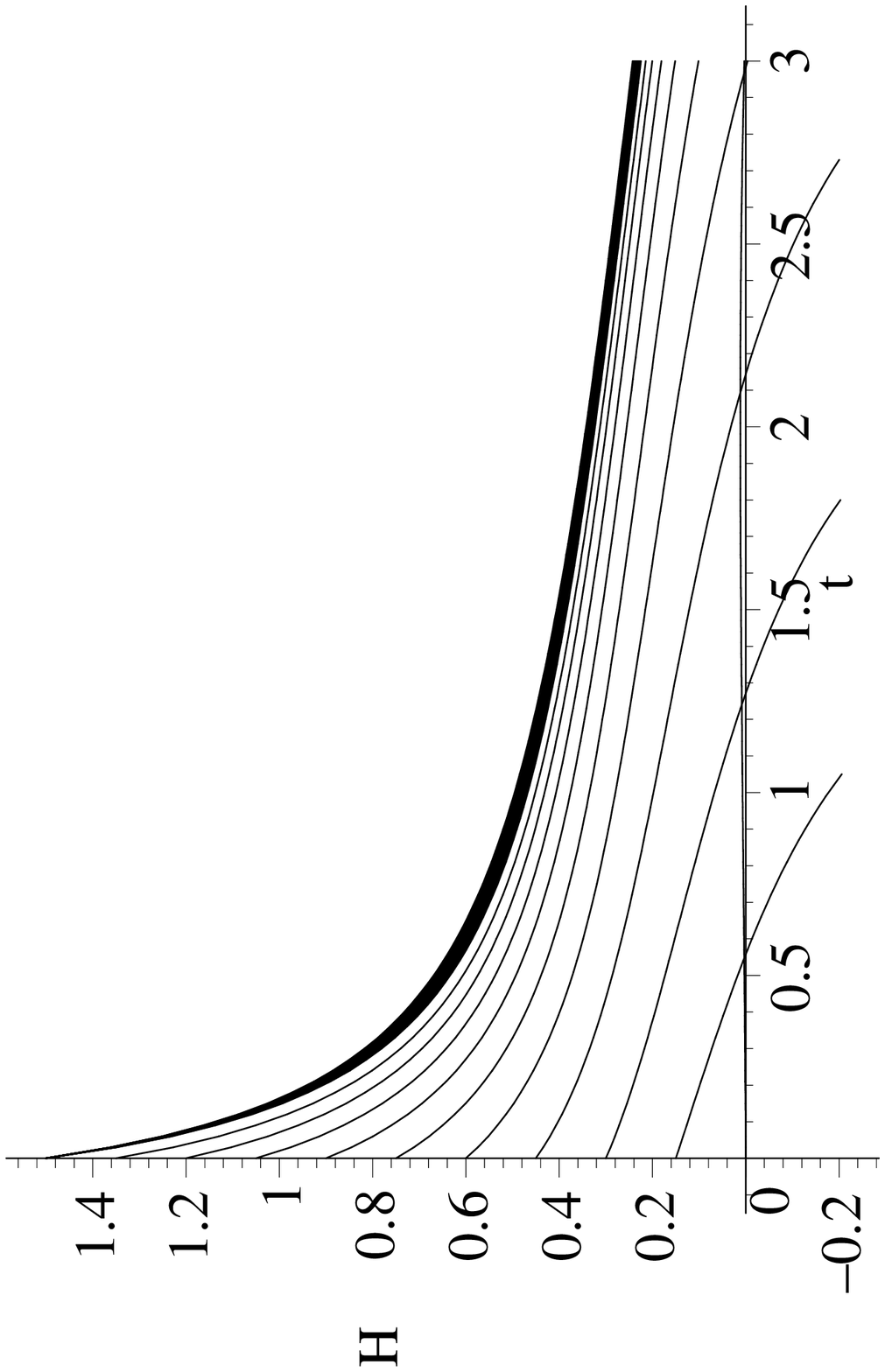}{Evolution of the Hubble constant $H$ with
parameters as in Fig. \ref{he6}.}{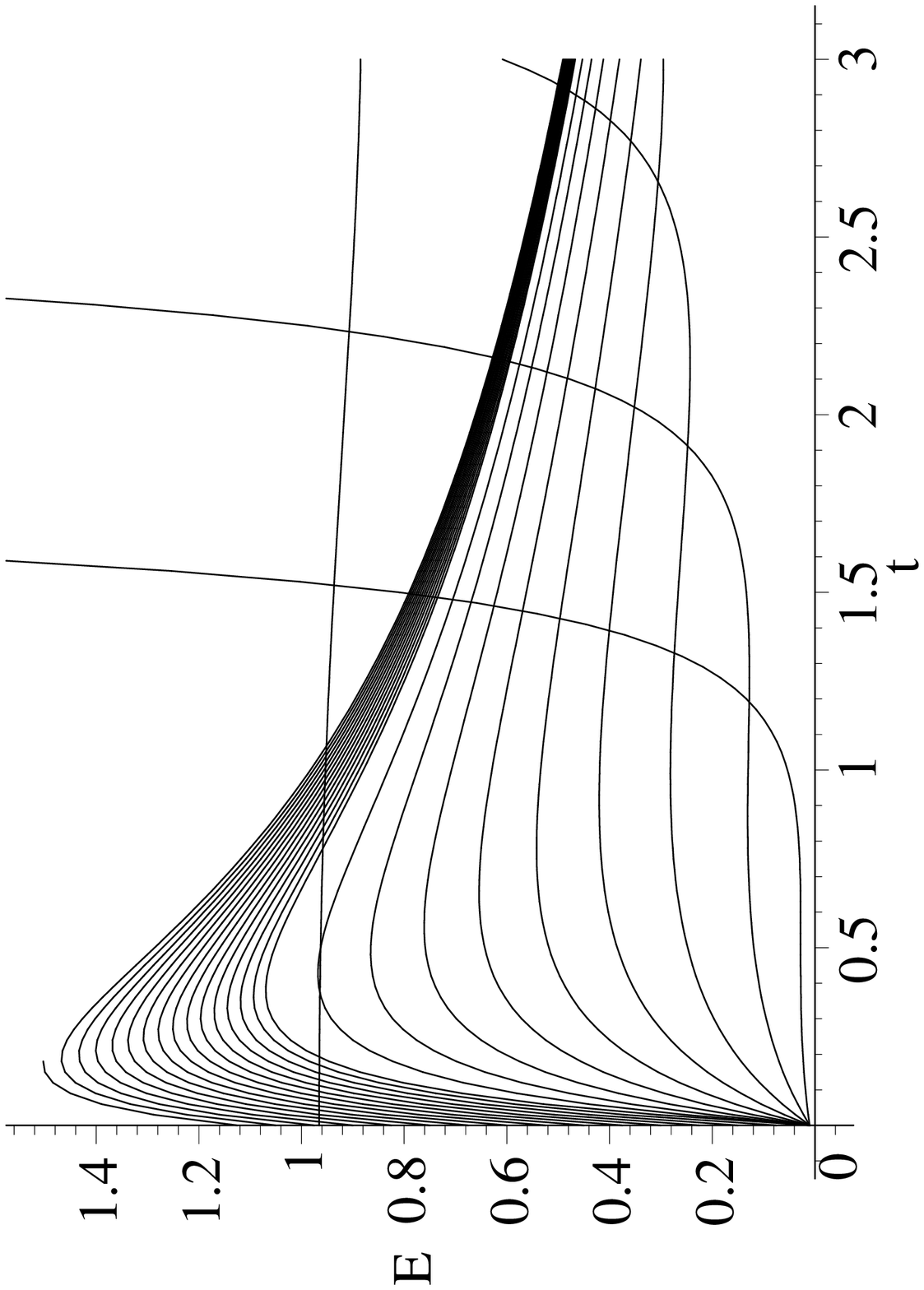}{Evolution of the energy
density $\ve$ with parameters as in Fig. \ref{he6}.}{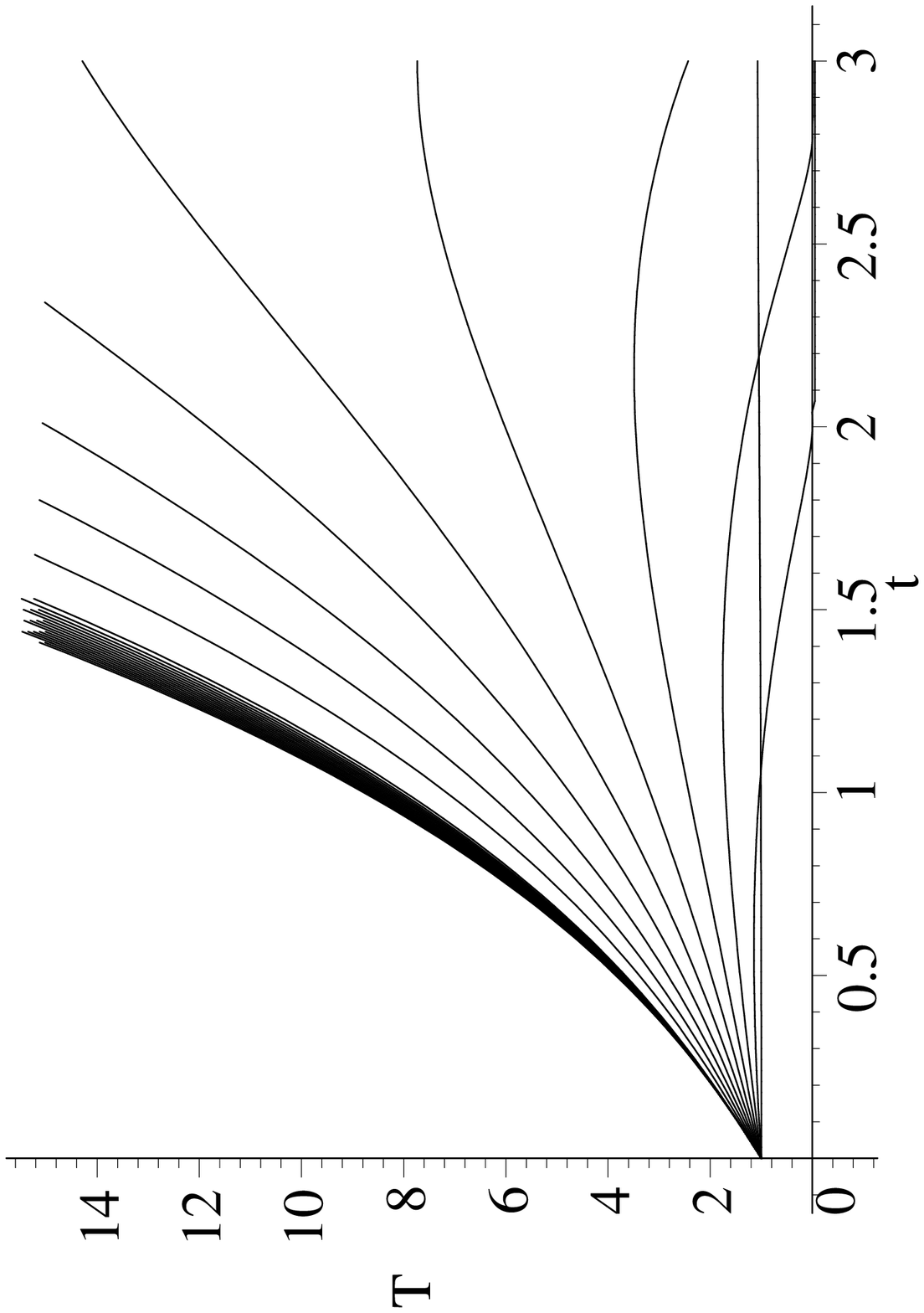}{Evolution
of the volume scale $\tau$ with parameters as in Fig. \ref{he6}.}

\myf{0.20}{0.27}{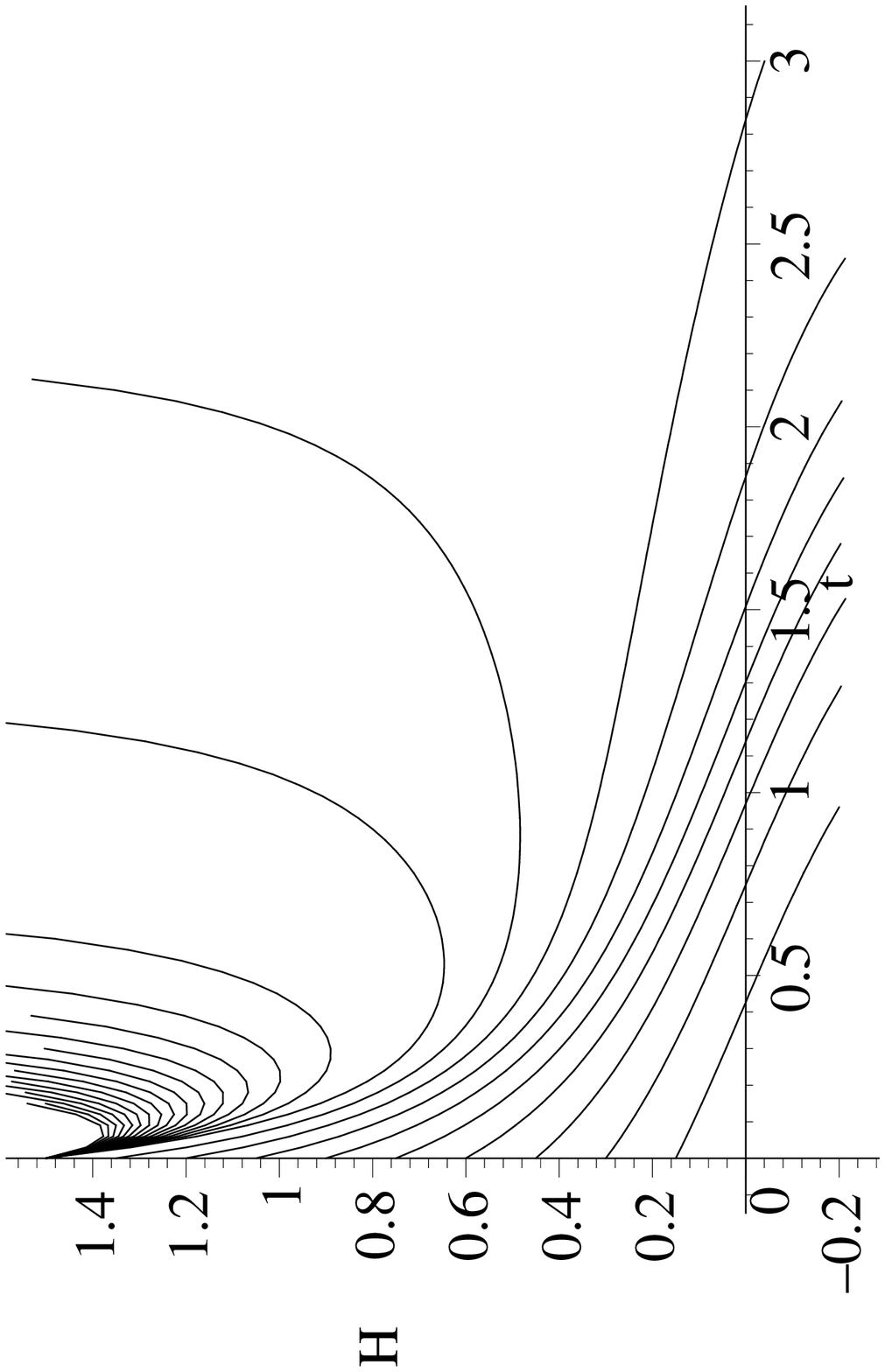}{Evolution of the Hubble constant $H$ with
parameters as in Fig. \ref{he7}.}{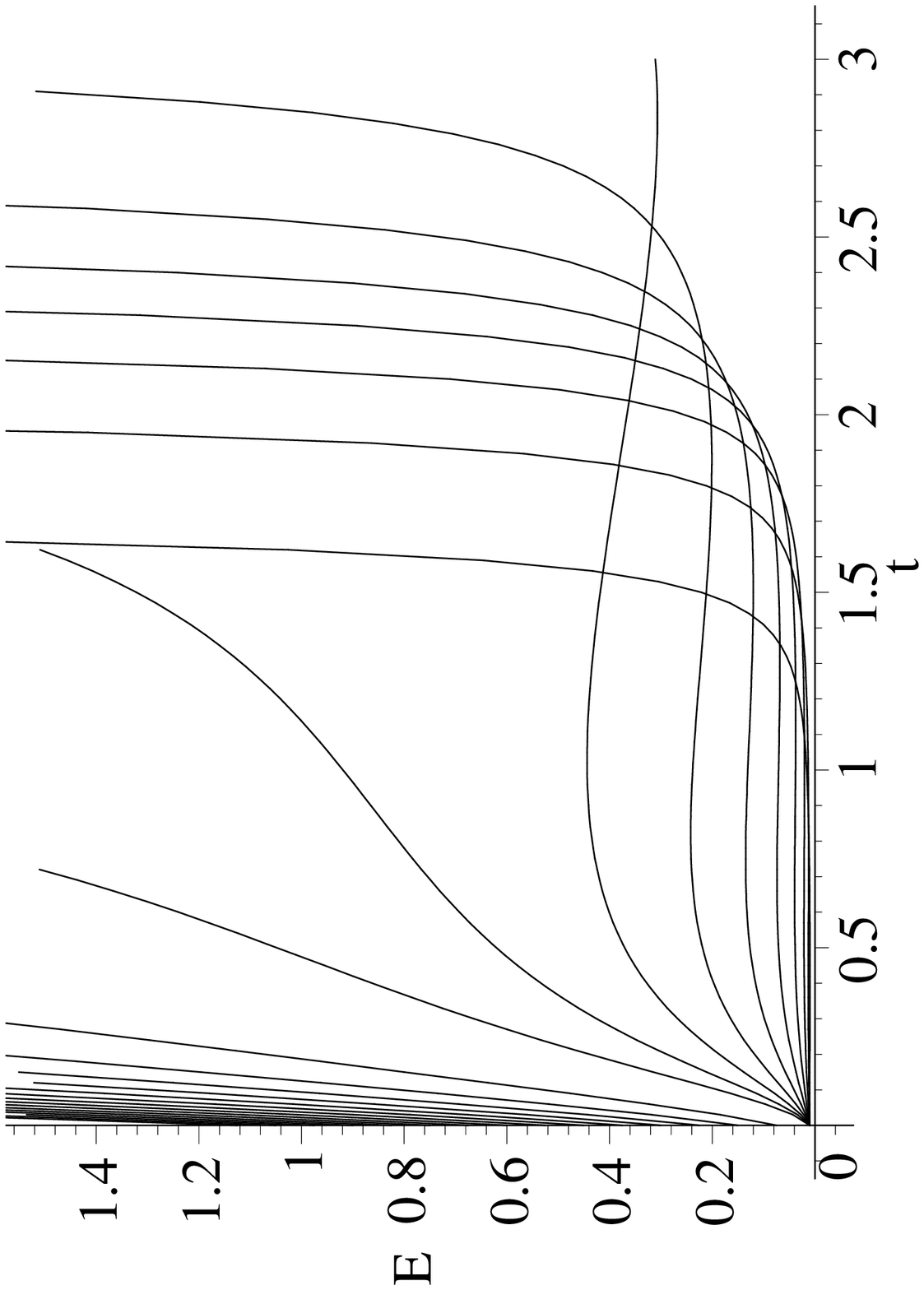}{Evolution of the energy
density $\ve$ with parameters as in Fig. \ref{he7}.}{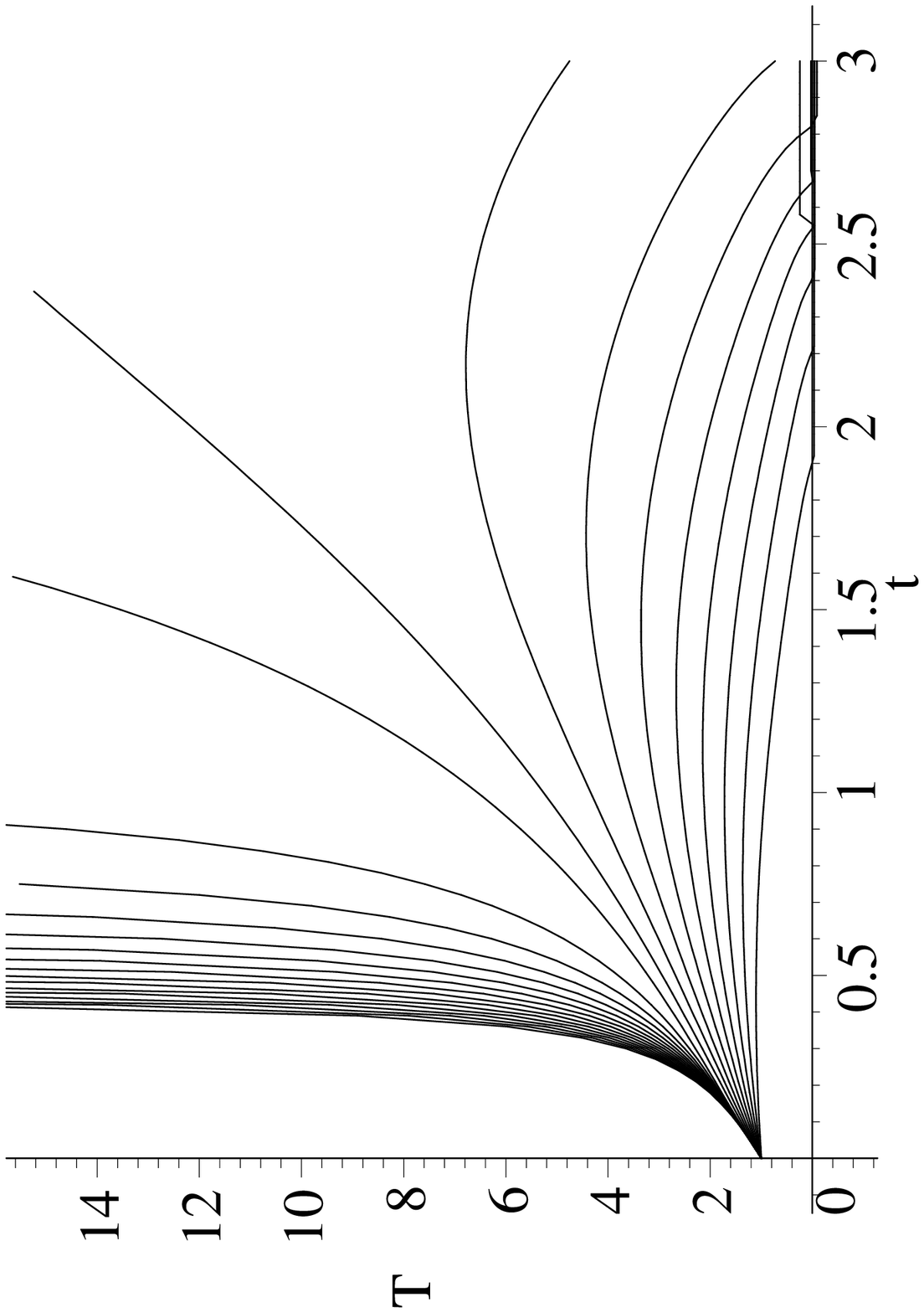}{Evolution
of the volume scale $\tau$ with parameters as in Fig. \ref{he7}.}

\end{document}